\documentclass[final,3p,times,twocolumn]{elsarticle}

\usepackage{amssymb}
\usepackage{graphicx}

\usepackage{amsmath} %Added by Author
\biboptions{sort&compress}
\usepackage{caption}
\usepackage{subcaption}
\usepackage[dvipsnames]{xcolor}
\usepackage{bm}
\usepackage{xfrac}
\usepackage{tikz,siunitx}
\usepackage{hyperref}
\usepackage{breqn}
\usepackage[numbers]{natbib}

\DeclareMathAlphabet{\mathcal}{OMS}{cmsy}{m}{n}

\newcommand{\tikzdash}[1]{\protect\tikz[baseline=-.65ex,thick]{\protect\draw[line width=1.5pt,dash pattern=on 3.5pt off 3.5pt] (0,0) -- (#1,0);}}

\newcommand{\tikzdot}[1]{\protect\tikz[baseline=-.65ex,thick]{\protect\draw[dotted] (0,0) -- (#1,0);}}

\newcommand{\tikzblue}[1]{\protect\tikz[baseline=-.65ex,thick]{\protect\draw[line width=1.5pt, color=blue] (0,0) -- (#1,0);}}

\newcommand{\tikzred}[1]{\protect\tikz[baseline=-.65ex,thick]{\protect\draw[line width=1.5pt, color=red] (0,0) -- (#1,0);}}

\newcommand{\tikzgreen}[1]{\protect\tikz[baseline=-.65ex,thick]{\protect\draw[line width=1.5pt, color=ForestGreen] (0,0) -- (#1,0);}}

\newcommand{\tikzpurple}[1]{\protect\tikz[baseline=-.65ex,thick]{\protect\draw[line width=1.5pt, color=orange] (0,0) -- (#1,0);}}

\newcommand{\tikzsim}[1]{\protect\tikz[baseline=-.65ex,thick]{\protect\draw[line width=1pt, color=Cerulean] (0,0) -- (#1,0);}}

\newcommand{\tikzpred}[1]{\protect\tikz[baseline=-.65ex,thick]{\protect\draw[line width=1.5pt,dash pattern=on 3pt off 3pt, color = Peach] (0,0) -- (#1,0);}}

\journal{Journal of Sound and Vibration}

\begin{document}

%\mbox{}
%\nomenclature{\(c\)}{Speed of light in a vacuum}
%\nomenclature{\(h\)}{Planck constant}

\begin{frontmatter}



\title{On Machine Learning-Driven Surrogates for Sound Transmission Loss Simulations}

\author[inst1,inst2]{Barbara Zaparoli Cunha}
\author[inst4]{Abdel-Malek Zine}
\author[inst1]{Mohamed Ichchou}\corref{cor1}
\author[inst3]{Christophe Droz}
\author[inst2]{Stéphane Foulard}

\cortext[cor1]{Corresponding author at École Centrale de Lyon.
36, Avenue Guy de Collongue, 69134, Écully, France.
E-mail address: mohamed.ichchou@ec-lyon.fr (M. Ichchou)}

\affiliation[inst1]{organization={Laboratory of Tribology and Dynamics of Systems},%Department and Organization
            addressline={Ecole Centrale Lyon}, 
            city={Ecully},
            postcode={69130}, 
            %state={Rhône},
            country={France}}

\affiliation[inst2]{organization={Compredict GmbH},%Department and Organization
            addressline={}, 
            city={Darmstadt},
            %postcode={64283}, 
            %state={Hessen},
            country={Germany}}
            
\affiliation[inst3]{organization={Univ. Gustave Eiffel, Inria},%Department and Organization
            addressline={COSYS/SII, I4S team}, 
            city={Rennes},
            %postcode={35042}, 
            %state={Brittany},
            country={France}}
            
\affiliation[inst4]{organization={Institut Camille Jordan},%Department and Organization
            addressline={Ecole Centrale Lyon}, 
            city={Ecully},
            %postcode={69130}, 
            %state={Rhône},
            country={France}}

\begin{abstract}
    Surrogate models are data-based approximations of computationally expensive simulations that enable efficient exploration of the model's design space and informed decision-making in many physical domains.
    The usage of surrogate models in the vibroacoustic domain, however, is challenging due to the non-smooth, complex behavior of wave phenomena.
    This paper investigates four Machine Learning (ML) approaches in the modelling of surrogates of Sound Transmission Loss (STL).
    Feature importance and feature engineering are used to improve the models' accuracy while increasing their interpretability and physical consistency.
    The transfer of the proposed techniques to other problems in the vibroacoustic domain and possible limitations of the models are discussed.

\end{abstract}

% \includegraphics[width=0.95\textwidth]{Figures/Graphical_Abstract.pdf}

% \begin{highlights}
% \item The benchmarking of ML-based surrogate models for the STL analyses was carried out.

% \item STL models with different methods of complexity are analyzed.

% \item Sensitivity analysis allows checking the surrogate physical consistency.

% \item NN-based surrogates with physics-guided features consistently performed better.

% \end{highlights}


\end{frontmatter}

% \linenumbers

\section{Introduction}
\label{sec:Introduction}

A thorough analysis of how the system's parameters influence its response is crucial for an optimal and robust product design, which reduces time and costs in the late stages of product development. 
Running high-fidelity simulations of the physical system, however, requires a large amount of computational resources, and, thus, extensive exploration over the design space is prohibitive.
Because of this, surrogate models have been used in different domains as efficient tools for decision-making and risk management \citep{sobester2008engineering, queipo2005surrogate, bhosekar2018advances}. 
Surrogates or metamodels are data-driven methods that extract and learn information from input-output pairs obtained from a physics-based simulation to create a proxy with short evaluation time and little loss of accuracy.

Surrogate models have been applied to the vibroacoustic domain as Noise, Vibration and Harshness (NVH) performance arises as a key indicator of customer satisfaction and vibroacoustic simulations are computationally costly due to, e.g., the complex behavior involved in fluid-structure interactions.
In many studies, the system response is approximated by a polynomial using the Response Surface Methodology (RSM) \citep{wang2017structural,liang2007acoustic,lu2017design,azadi2009nvh,guo2022research,chen2020investigation}.
Second-order polynomials are usually used in RSM for their sample efficiency and interpretability, but they are unable to capture arbitrary nonlinearities in the system response. 
On the other hand, higher-order polynomials are rarely employed due to the costly training \citep{moustapha2016adaptive}.
Therefore, machine learning-based surrogates have been used in the vibroacoustic analysis for their capability of learning complex data relations and non-linear functions and applied in optimization \citep{kiani2016comparative, cha2004optimal, bacigalupo2020machine} and uncertainty analyses \citep{diestmann2021surrogate, Chai2020, Nobari2015, HURTADO2001113, Cicirello2020}.

A surrogate model using Gaussian Process Regressor (GPR) was implemented in \citep{cha2004optimal} to optimize the STL of an intake system. \citet{casaburo2021gaussian} trained a GPR-based surrogate model to predict the increase of the STL response at resonance frequencies of porous acoustic meta-materials.
The results showed that better accuracy can be achieved when a more feature-rich STL model is used.
\citet{bacigalupo2020machine} performed the optimization of the band-gap of acoustic metamaterials using Radial Basis Function (RBF)-based surrogate. 
Surrogates for crash and NVH response of a Body-In-White model using RBF and Neural Network (NN)-based algorithms were designed in \citep{kiani2016comparative} and \citep{li2021vehicle}, respectively. 
\citet{ibrahim2020surrogate} investigates various ML-based surrogate models to predict the noise radiated by electric motors and concluded that global models encompassing structural, electromagnetic, and acoustic physical domains performed better than local surrogates.
NN-based surrogate models have also accurately replaced boundary element methods to predict interior vehicle sound pressure level \citep{zhang2018virtual}.

The growing trend in vibroacoustic of relying on surrogates based on high-fidelity simulations not just enables the search for optimal and reliable designs, but also paves the way to the construction of Digital Twins for real-time vibroacoustic applications, such as online monitoring and control \citep{Barkanyi2021,cunha2022review}.
However, the reliability of these solutions depends on the accuracy of the surrogate models.
In particular, in vibroacoustic, the typical non-smooth functions pose difficulty to the construction of accurate ML-based surrogates as ML methods assume local smoothness of data \citep{Mehta2019, domingos2012few}.
Therefore, to ensure that further surrogate-based developments on vibroacoustic are well-founded, it is imperative to develop scientific-based guidelines to construct physically consistent and accurate ML-based surrogates considering the particulars of the domain.
Thus, the objective of this paper is to identify ML methods suitable to model vibroacoustic problems, as well as techniques to incorporate domain knowledge into the surrogate. Furthermore, this work investigates the application of sensitivity analysis methods embedded in the ML to improve surrogate interpretability and enable reasoning about the physical phenomena. These investigations are carried out for the classical acoustic problem of STL formulated with different complexity levels, as the STL problem encompasses important vibroacoustic phenomena well understood by the acoustic community \citep{santoni2020review}.

Section \ref{sec:TL} describes the methodology followed to construct four different physics-based models of STL. 
The workflow of the surrogate implementation is presented in Section \ref{sec:Surrogate}. In Section \ref{sec:Feature Engineering},  the sensitivity analyses results and the effects of adding physics-guided features are shown. 
Subsequently, in Section \ref{sec:Benchmarking}, a benchmarking analysis compares the performance of each ML method used to construct the surrogate for different numbers of supporting points and levels of complexity of the STL problem. 
The pros and cons of each ML method are discussed. The influence of design space size and location is also investigated in Section \ref{sec:Benchmarking}. Section \ref{sec:Conclusions} presents the paper's conclusion and a discussion of the transferability of its findings to other vibroacoustic problems.

The data and algorithms used in this work are available at \href{https://github.com/ZaparoliCunha/STL_Surrogate.git}{GitHub}.

%%%%%%%%%%%%%%%%%%%%%%%%%%%%%%%%%%%%%%%%%%%%%%%%
%%%%%%%%%%%%%%%%%%%%%%%%%%%%%%%%%%%%%%%%%%%%%%%%
% Methodology
%%%%%%%%%%%%%%%%%%%%%%%%%%%%%%%%%%%%%%%%%%%%%%%%
%%%%%%%%%%%%%%%%%%%%%%%%%%%%%%%%%%%%%%%%%%%%%%%%

\section{Physics-Driven Models of Sound Transmission Loss}\label{sec:TL}

\begin{figure} [h]
\centering 
\includegraphics[width=0.42\textwidth]{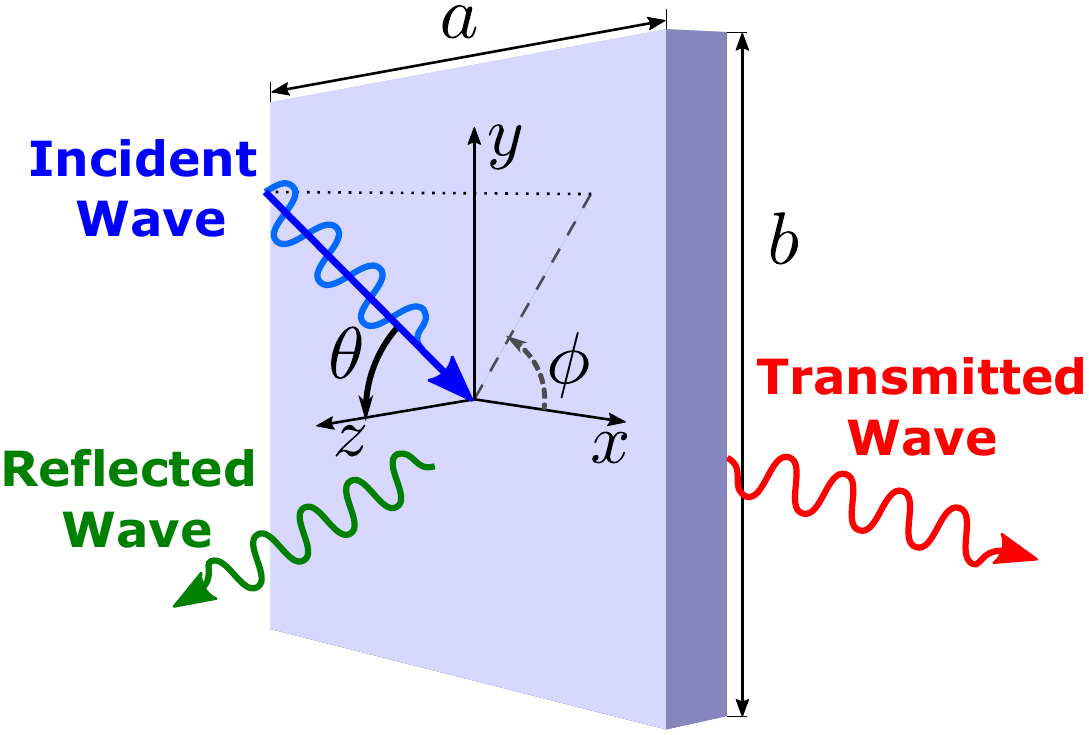} 
\caption{\label{fig: STL} Illustration of the Sound Transmission Loss (STL) analysis of a plate.} 
\end{figure}

When an acoustic wave hits a plate, it is partially reflected in the incident acoustic field and partially absorbed by the plate itself.
Most of the absorbed portion excites the plate, making it vibrate and radiate noise in both the incident and transmitted fields,  as illustrated in Figure \ref{fig: STL}. 
The Sound Transmission Loss (STL) is defined by the ratio between the incident sound power $P_I$ and the sound power transmitted through the plate $P_T$, being usually evaluated in decibels as

\begin{equation}
    STL = 10 \log_{10} {\left|\frac{P_I}{P_T}\right|^2}.
\end{equation}

The transmitted noise is inversely proportional to the system effective impedance, which is the sum of the mechanical impedance of the plate and the fluid-loading impedance \citep{fahy2007sound}. 
The effective impedance depends on the fluid properties, the incident $\theta$ and azimuth $\phi$ angles of the incident wave, and the material, dimensions, and boundary conditions of the plate.
This paper considers that the incident field is a random field of incident plane waves in all directions, known as diffuse field.

At frequencies below the first natural frequency of the plate, the plate dynamics influence on the STL is negligible, and the STL is controlled by the plate bending stiffness,

\begin{equation}
    \label{eqn:bending_stiffness}
         D = \frac{E h^3}{12(1-\nu^2)}(1+i\eta),
    \end{equation}
    
\noindent where $h$ is the thickness, $E$ is the elastic modulus, $\nu$ is the Poisson's ratio and $\eta$ is the damping loss factor of the plate.

When the driving frequency is close to the first natural frequencies of the plate, the resonant behavior of the plate governs the STL. As the frequency increases and the resonant behavior attenuates, the plate behavior approaches that of a limp mass, especially for thin plates in frequencies where the bending stiffness is not relevant. At this frequency range, the STL depends solely on the surface mass density $m=h\rho$ and can be approximated by the mass law, as demonstrated in \citep{fahy2007sound}.

\begin{figure} 
\centering 
\includegraphics[width=0.45\textwidth]{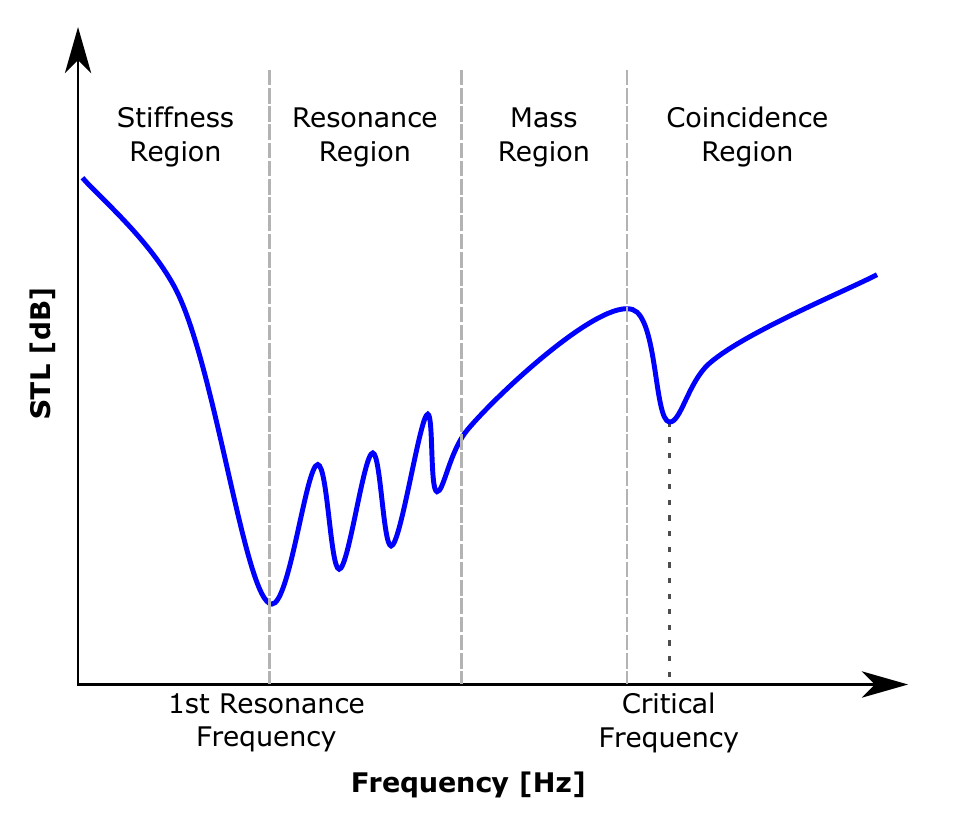}
\caption{\label{fig: STL_curve}Typical Sound Transmission Loss (STL) curve as a function of frequency, showing the stiffness, resonance, mass and coincidence regions.}
\end{figure}

With the increase in frequency, the coincidence phenomenon takes place, and the mass law is no longer valid. The coincidence happens when the trace velocity of the acoustic wave projected in the plate is equal to the velocity of its natural bending wave. These waves superpose and create a scenario of minimal impedance, maximizing the transmission of sound. The coincidence frequency $\omega_{coinc}$ in which the trace and the bending wavenumbers coincide is evaluated by:

\begin{equation}
\label{eqn:coincidence_frequency}
    \omega_{coinc} = \frac{c_0^2}{\sin^2{\theta}}\sqrt{\frac{\rho h}{D}},
\end{equation}
where $c_0$ is the characteristic sound speed, $\rho$ is the plate density. The minimum frequency in which coincidence occurs is for grazing incidence ($\theta = 90^{\circ}$) and is called critical frequency $\omega_{crit}$, where the amplitude of the valley in the STL is controlled by the plate damping. Above the critical frequency, the coincidence phenomenon occurs for gradually smaller incidence angles and the STL is controlled by the plate bending behavior. Figure \ref{fig: STL_curve} illustrates a typical STL curve and its influence regions.

In the literature, different STL models capture the aforementioned curve behaviors with distinct levels of detail \citep{santoni2020review}. In this paper, four STL methodologies are addressed to evaluate the ML-based surrogate capabilities. The first and simplest methodology considered is the analytical STL solution for an \emph{infinite plate} proposed in \citep{cremer1942theorie} and implemented in accordance to \citep{christen2016wave}. As the unbounded infinite plate is not affected by the resonances, the STL in low-frequencies relies only on the surface density mass.

To account for the plate width $a$ and length $b$, the radiation efficiency of a finite plate may be considered instead of the radiation efficiency of the infinite plate. Spatial windowing and Rayleigh-integral-based techniques can be used for this purpose, resulting in a correction factor applied to the infinite transmission transparency \citep{santoni2020review}. The \emph{correction factor} approach implemented in \citep{rhazi2010simple} modifies mainly the results below the critical frequency, where the plate modal behavior determines the sound radiated. Although it is a simplified approach, the modified STL curve is coherent with experimental curves with band-average \citep{santoni2020review}.

Detailed STL evaluation of bounded plates must account for its structural response under the forced vibrations. The plate modes' natural frequencies and radiation efficiencies control the STL, resulting in a highly rough behavior. The current work implements \emph{Modal Summation} (MS) and \emph{Finite Element Method} (FEM) to fully model the vibroacoustic phenomena in the STL of a finite plate with simply supported boundary conditions. The MS solution approximates the plate displacement as the sum of the contribution of the first modal shapes, which is analytically evaluated as described in \citep{xin2010sound, wang2015modal}. The FEM model, implemented in COMSOL Multiphysics\textsuperscript{\textregistered} software \citep{COMSOL_soft}, solves the structural and acoustic field equations numerically by modeling plate and the transmitted field as finite elements. The diffuse field is modeled as a harmonic pressure with components in random directions which applies to the plate surface.

In practical scenarios, only the overall STL behavior defines the plate design and, thus, the band-averaged STL can be considered. Therefore, the results obtained with MS and FEM are also evaluated in one-third octave frequency bands. In summary, six different physics-driven models of STL serve as ground truth for the surrogate models:
\begin{itemize}
\setlength\itemsep{-0.1em}
    \item analytical solution for infinite plate,
    \item finite plate STL approximated by correction factor,
    \item analytical Modal Summation (MS) approach,
    \item MS approach with band-average,
    \item numerical FEM,
    \item FEM with band-average.
\end{itemize}

For further details on the methodologies of STL analyses implemented, refer to~\ref{sec:sample:appendix1}.

\section{Development of the Machine Learning-Driven Surrogate Models}\label{sec:Surrogate}

\begin{figure}[ht] 
\centering 
\includegraphics[width=0.3\textwidth]{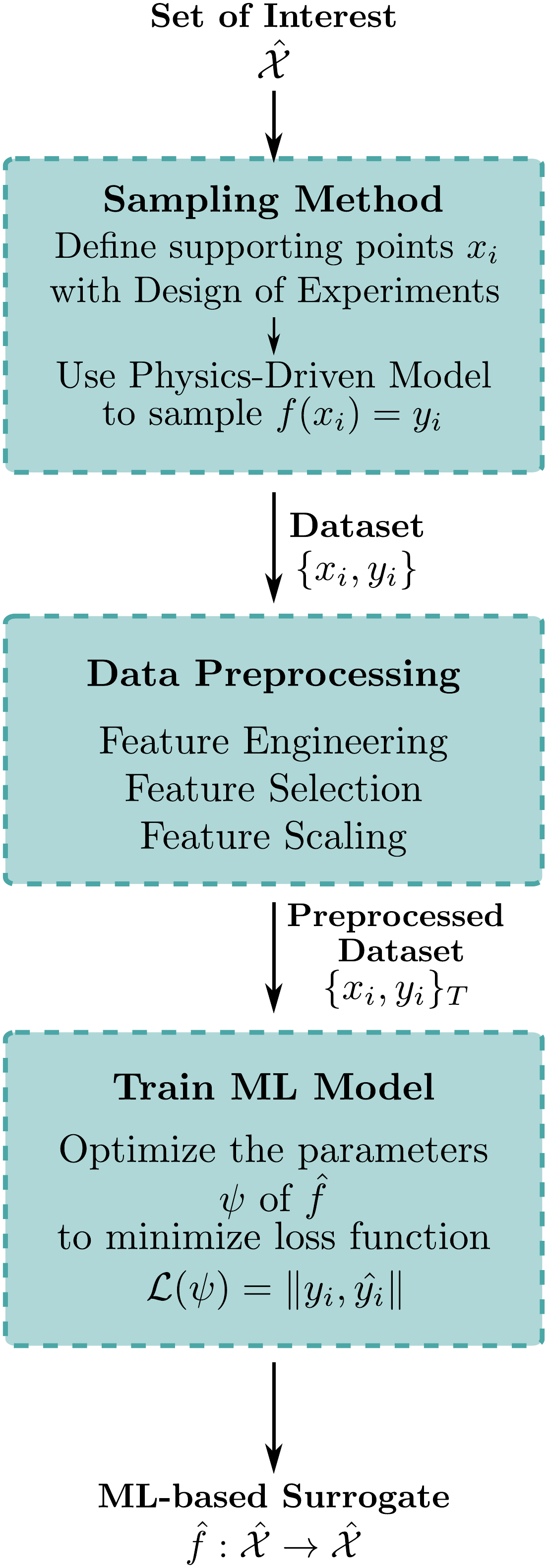}
\caption{\label{fig: surrogate}Fundamental workflow of Machine Learning (ML)-based surrogate models.} 
\end{figure}

A surrogate model is a computationally inexpensive mathematical approximation of a complex simulation.
More formally, let $f:  \mathcal{X} \rightarrow  \mathcal{Y}$ be the simulated function.
The surrogate model can be defined as a function $\hat{f}: \hat{\mathcal{X}} \rightarrow \hat{\mathcal{Y}}$, where $\hat{\mathcal{X}} \subset  \mathcal{X}$ is the set of interest for the surrogate model, and $\hat{\mathcal{Y}}$ is the associated codomain.
The relation between the simulated function and the surrogate is given by $f(x_i) = y_i = \hat{f}(x_i) + \epsilon_i,\; \forall x_i \in \hat{X}$, where $\epsilon_i$ is the approximation error.
The predictor $\hat{f}(x)$ is fitted with $N$ input/output pairs $\{x_i, y_i\}$ sampled from the simulated function to minimize the error $\epsilon_i$ for all inputs in the set of interest.
Because the number of samples $N$ must be kept small due to the high computational cost of evaluating $f(x)$, choosing an efficient sampling method is crucial for designing an accurate surrogate model.
In the following, the basic workflow for creating a surrogate model is described, and Figure \ref{fig: surrogate} shows a block diagram with the main steps.

\textbf{\emph{Sampling methods.}}
The first stage in designing a surrogate model is to define informative supporting points using an adequate sampling strategy. 
In random sampling, all points in the interval of interest have an equal probability of being drawn.
Although having low bias and being easy to deploy, 
it is known that random sampling can lead to gaps in the input domain. 
Thus, another strategy is to segment the input space into $M$ intervals along each dimension and randomly sample from each interval.
This method is the so-called Latin Hypercube Sampling (LHS) and promises to produce a better depiction of the input domain of the target function \citep{dwight2012tutorial}. Therefore, this paper used LHS to define the supporting points for the training of the surrogates. 

\textbf{\emph{Data Preprocessing.}}
The accuracy of data-driven models can improve significantly with the proper operations over the feature set.
Here, features refer to the different dimensions of the input data $x \in  \mathcal{X}$.
In feature engineering, domain knowledge is used to select the most informative features of an input set and to add physics-guided features through basic operations over the raw data to improve the predictor's accuracy and interpretability.
Feature scaling is another crucial preprocessing step, especially for ML predictors which are distance-based,  e.g., GPR, or gradient-based ones, such as NN.
The goal of feature scaling is to remove the effect of the different range of values in the input so that equal importance is given to each feature in the fitting of the regressor.
Also, it guarantees that the gradient descent is updated with similar rates for all features, improving convergence.
Normalization and standardization are the main methods for feature scaling.

\textbf{\emph{ML-driven Surrogate Models.}}
The predictor function $\hat{f}(x)$ can be fitted with different ML models.
In this paper, four of the most used models for surrogate modeling are considered, namely Neural-Networks (NN), Gaussian Process Regressor (GPR), Random Forest (RF), and Gradient Boosting Trees (GBT).
For more details on the approaches, refer, e.g., to \citep{murphy2012machine}.
In NN, input data passes through a pipeline of linear transformations and non-linear functions.
Accuracy is usually measured by the Euclidean distance between the predictor output $\hat{f}(x_i)$ and the target value $y_i$, and the model is fitted with gradient descent via backpropagation.
GPR uses Bayes' theorem to update the prior assumption over the model parameters with the sampled input-output pairs and is well-suited for regression with few samples available \citep{rasmussen2003gaussian}.
The regressor output is the probabilistic model that best describes the correlation of all points in the domain.
Both RF and GBT build upon the decision tree learning paradigm.
RF is an ensemble model made of several decision trees to reduce the variance of single tree predictions, which is achieved by averaging the output of each decision tree in the RF. 
In GBT, decision trees are trained sequentially where each decision tree trains on the mispredicted data of the previous one, thus, focusing mainly on reducing the predictor bias.

\bigskip

After the surrogate is trained, it can predict the system response in new points of the design space and be applied in optimization and uncertainty propagation of the system parameters. Furthermore, sensitivity analyses can be used to increase the black-box interpretability and can either be obtained as a by-product of ML methods or be straightforwardly implemented \citep{sudret2017surrogate, cao2016advanced, pizarroso2020neuralsens, casalicchio2018visualizing}. 
The Mean Decrease in Impurity (MDI) \citep{louppe2013understanding} is obtained after the training of the RF model, and it is applied for sensitivity analysis as it approximates the normalized total Sobol's indices \citep{jaxa2018tree}. 
Further discussion and methods for interpreting ML methods are presented in \citep{molnar2020interpretable}.
In Section \ref{sec:Feature Engineering} this method was applied to verify the surrogate physical consistency. 

%%%%%%%%%%%%%%%%%%%%%%%%%%%%%%%%%%%%%%%%%%%%%%%%%%%%%%%%%%%%%%%%%%%%%%%%%%%%%%%%%%%
% Sensitivity Analysis
%%%%%%%%%%%%%%%%%%%%%%%%%%%%%%%%%%%%%%%%%%%%%%%%%%%%%%%%%%%%%%%%%%%%%%%%%%%%%%%%%%

\section{Sensitivity Analysis of Random Forest Surrogate and Influence of Physics-Guided Features}\label{sec:Feature Engineering}

\begin{figure} [h]
\centering 
\includegraphics[width=0.35\textwidth]{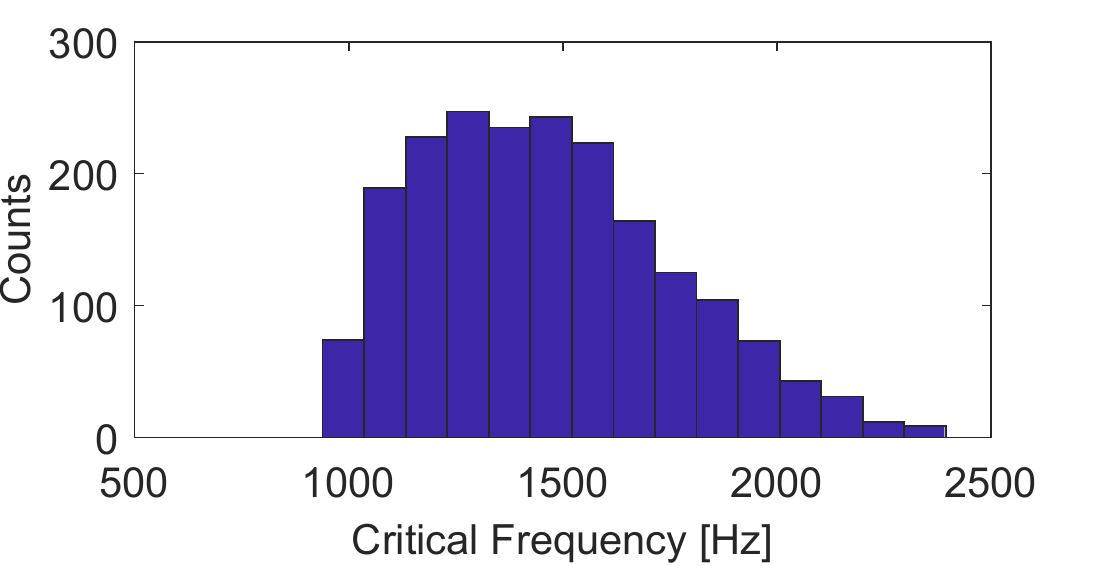}
\caption{\label{fig:Range_CriticalFrequency} Histogram of the critical frequencies from the STL designs in the set of interest from Table \ref{tab:SDS}.} 
\end{figure}

In this section, the consistency of ML-based surrogates with physics-driven models of STL is investigated.
Sensitivity analysis is used to demonstrate the importance of each parameter in the system response.
For exemplary purposes, results are presented using MDI-based sensitivity indices of the RF-based surrogate as it is readily available after the training of the model. One RF regressor was trained for each frequency of interest to obtain the features importances for each output frequency
\footnote{The RF surrogate was implemented using \textit{sklearn} library in Python. The Multi-Output Regressor method of \textit{sklearn} was used to train one RF regressor for each output frequency. Each RF was trained with a maximum of 200 decision trees without a depth limit. The impurity criteria are the mean squared error.}.
The surrogates were trained with 2000 supporting points from the design space of Table \ref{tab:SDS} using LHS for the sampling. The critical frequencies for this design space range between approximately 1000 and 2500 Hz, as shown in Figure \ref{fig:Range_CriticalFrequency}.
The addition of two physics-guided features to the RF models, namely the mass density $m$ and the real part of the bending stiffness $D_R$, is evaluated.
According to STL theory, the mass density controls the STL in the mass region, while bending stiffness has a significant influence on both the stiffness and coincidence regions.
The Root Mean Squared Error (RMSE) is used to compare the accuracy of surrogates with and without physics-guided features as input.
A desktop with a hexa-core 3.1 GHz processor and 32 GB of RAM was used to run the training of the surrogates.

\begingroup 
    \setlength{\tabcolsep}{6pt} % Default value: 6pt 
    \renewcommand{\arraystretch}{1} % Default value: 1 
    \begin{table}[ht!] 
          \begin{center} 
            \caption{Design space with set of interest used to sample supporting points.} 
            \label{tab:SDS} 
            \begin{tabular}{c | c c c c c c c} 
              \bm{ \ } & \bm{$\rho$ } & \bm{$E$} & \bm{$\nu$} & \bm{$\eta$} & \bm{$h$} & \bm{$a$} & \bm{$b$}   \\ 
              {\ } & {$\sfrac{kg}{m^3}$ } & {$GPa$} & {$-$} & {$\%$} & {$mm$} & {$ m$} & {$m$}   \\ 
              \hline
              \textbf{min} & 2000 & 60 & 0.25 & 0.1 & 5 & 0.3 & 0.3   \\ 
              \textbf{max} & 3000 & 150 & 0.35 & 2.0 & 7 & 0.6 & 0.6   \\ 
            \end{tabular} 
          \end{center} 
    \end{table} 
\endgroup

\begin{figure} [h!]
\captionsetup[subfigure]{justification=centering}
\centering 
     \begin{subfigure}[b]{0.37\textwidth}
         \centering
         \includegraphics[width=\textwidth]{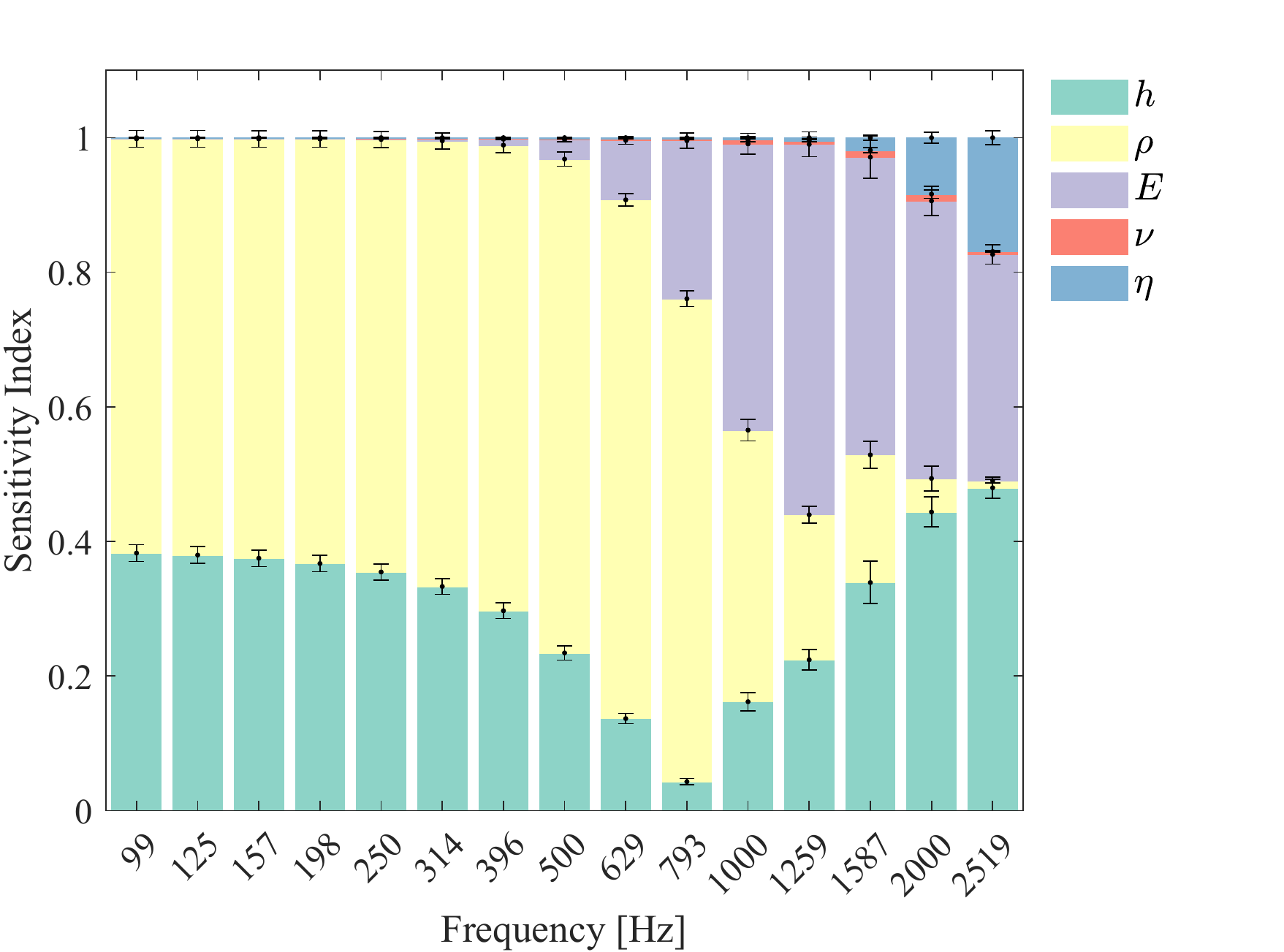}
         \caption{The inputs of the surrogate only consist of the STL variables. $RMSE = 0.21 \pm 0.00 \  dB$. Training time: 8.83s}
         \label{fig:FI_Infinite_BI}
     \end{subfigure}
     \hfill
     \begin{subfigure}[b]{0.37\textwidth}
         \centering
         \includegraphics[width=\textwidth]{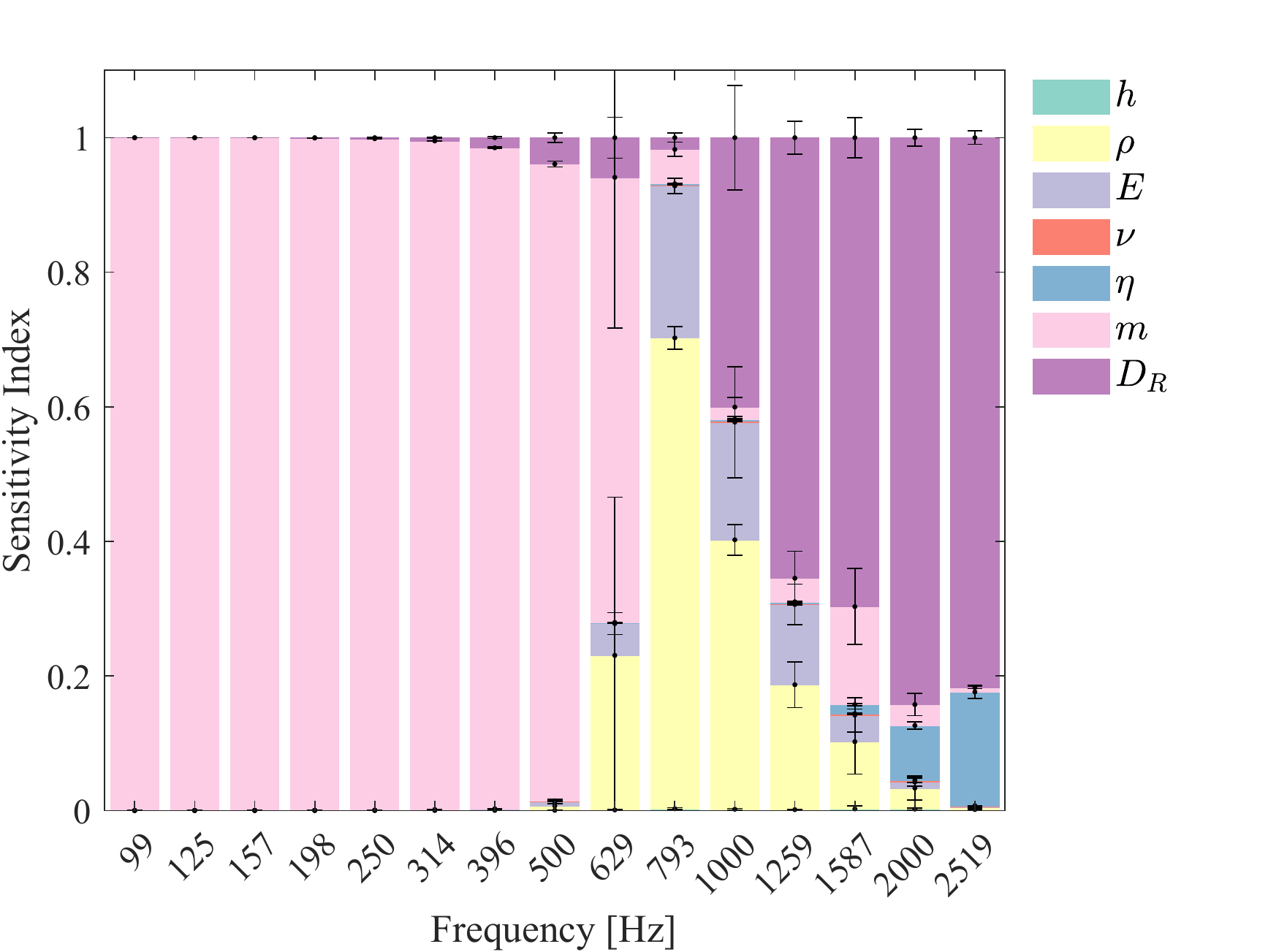}
         \caption{Physics-guided features $m$ and $D_R$ are included as input for the surrogate. $ RMSE = 0.12 \pm 0.01 \ dB$. Training time: 6.41s
         }
         \label{fig:FI_Infinite_EI}
     \end{subfigure}
        \caption{MDI-based sensitivity indices of the STL with the analytical infinite plate model.} 
        \label{fig:FI_Infinite}
\end{figure}

Figure \ref{fig:FI_Infinite_BI} and \ref{fig:FI_Infinite_EI} show the MDI-based sensitivity index obtained with the infinite plate models with and without the use of physics-guided features, respectively. The mass-controlled region is distinctly observable at low frequencies in both cases.
Approaching the coincidence region, the STL is controlled mainly by the plate density, and the thickness importance is negligible.
Despite the damping factor determining the STL amplitude at the critical frequency, \ref{fig:FI_Infinite_EI} shows that bending stiffness is the most significant feature at this frequency range. As noted in \citep{christen2016wave}, when a set of interest is considered rather than a specific plate design, the location of the critical frequency dip is more relevant for the STL result than its amplitude, which explains this result.
At higher frequencies in the coincidence region, the damping importance increases as for each frequency $\omega > \omega_{crit}$ there will be a component of the diffuse field with an angle $\theta$ such that $\omega_{coinc}(\theta)=\omega$ and the amplitude is again the major aspect of the STL. 
Finally, it is noted that the inclusion of the physics-guided features improves the surrogate accuracy from an RMSE of $0.21 \pm 0.00 \ dB$ to $0.12 \pm 0.01 \ dB$.

\begin{figure} [h!]
\captionsetup[subfigure]{justification=centering}
\centering 
     \begin{subfigure}[b]{0.37\textwidth}
         \centering
         \includegraphics[width=\textwidth]{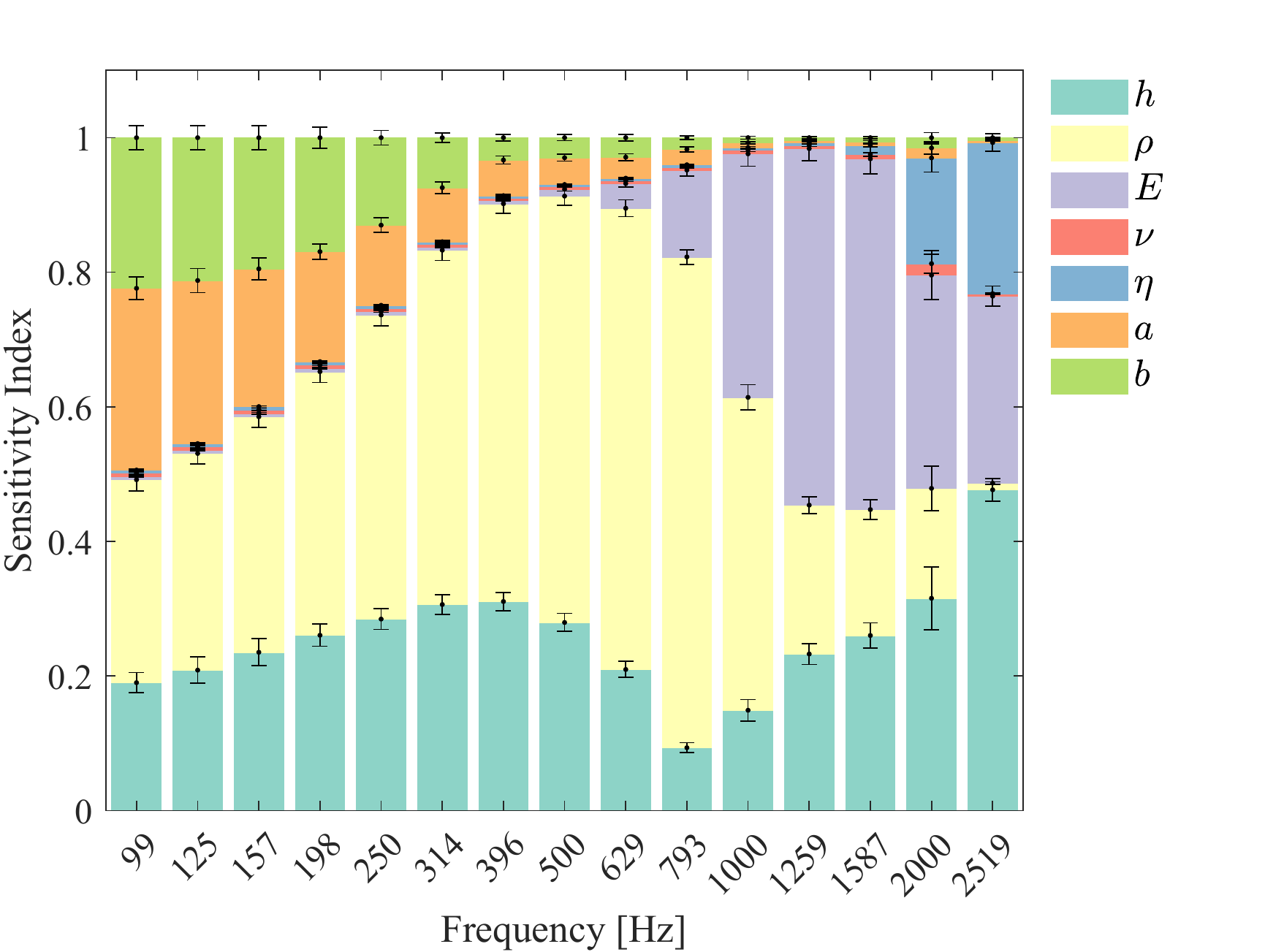}
         \caption{The inputs of the surrogate only consist of the STL variables. $RMSE = 0.36 \pm 0.01 dB$. Training time: 6.39s}
         \label{fig:FI_CR_BI}
     \end{subfigure}
     \hfill
     \begin{subfigure}[b]{0.37\textwidth}
         \centering
         \includegraphics[width=\textwidth]{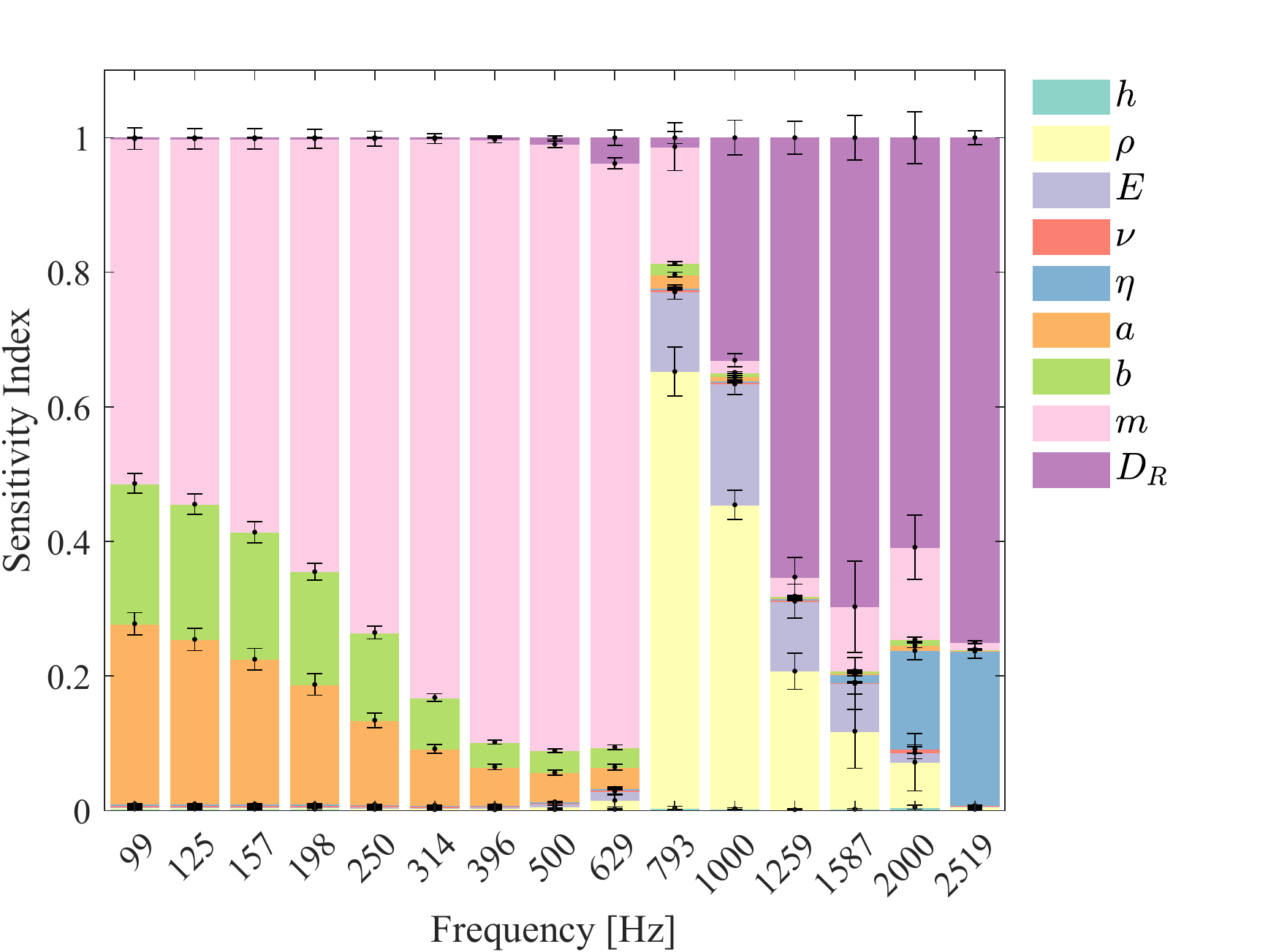}
         \caption{Physics-guided features $m$ and $D_R$ are included as input for the surrogate. $RMSE = 0.25 \pm 0.01 \ dB$. Training time: 13.29s}
         \label{fig:FI_CR_EI}
     \end{subfigure}
        \caption{MDI-based sensitivity of the STL of finite plates evaluated with the correction factor approach.} 
        \label{fig:FI_CR}
\end{figure}

The sensitivity indices of STL of finite plates based on the correction factor approach (Figure \ref{fig:FI_CR}) differ from the infinite plate indices only in the low-frequency region.
This result is coherent with the methodology, which applies a correction factor in the low-frequency range to account for the plate dimensions and comprehend its resonant behavior.
Although the plate stiffness would theoretically also impact the resonant modes of the structure, the STL evaluated by correction factor is insensitive to this feature in the resonant region of Figure \ref{fig:FI_CR}. Once more, the RMSE decreases from 0.36 dB to 0.25 dB with the inclusion of physics-guided features.

\begin{figure} [ht]
\captionsetup[subfigure]{justification=centering}
\centering 
     \begin{subfigure}[b]{0.37\textwidth}
         \centering
         \includegraphics[width=\textwidth]{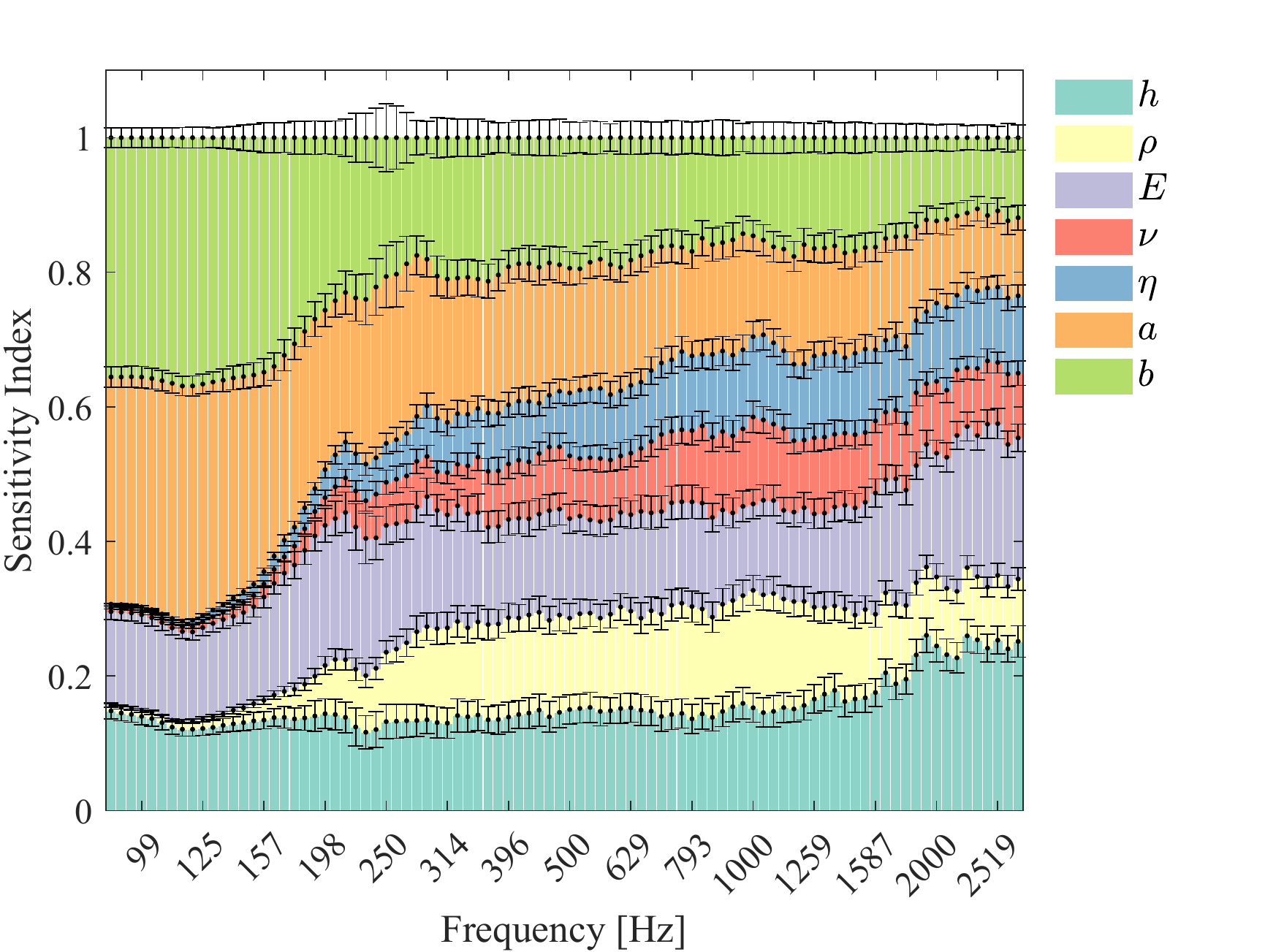}
         \caption{The inputs of the surrogate only consist of the STL variables. $RMSE = 3.38  \pm 0.03 \  dB$. Training time = 28.23s}
         \label{fig:FI_Modal_Summation_BI}
     \end{subfigure}
     \hfill
     \begin{subfigure}[b]{0.37\textwidth}
         \centering
         \includegraphics[width=\textwidth]{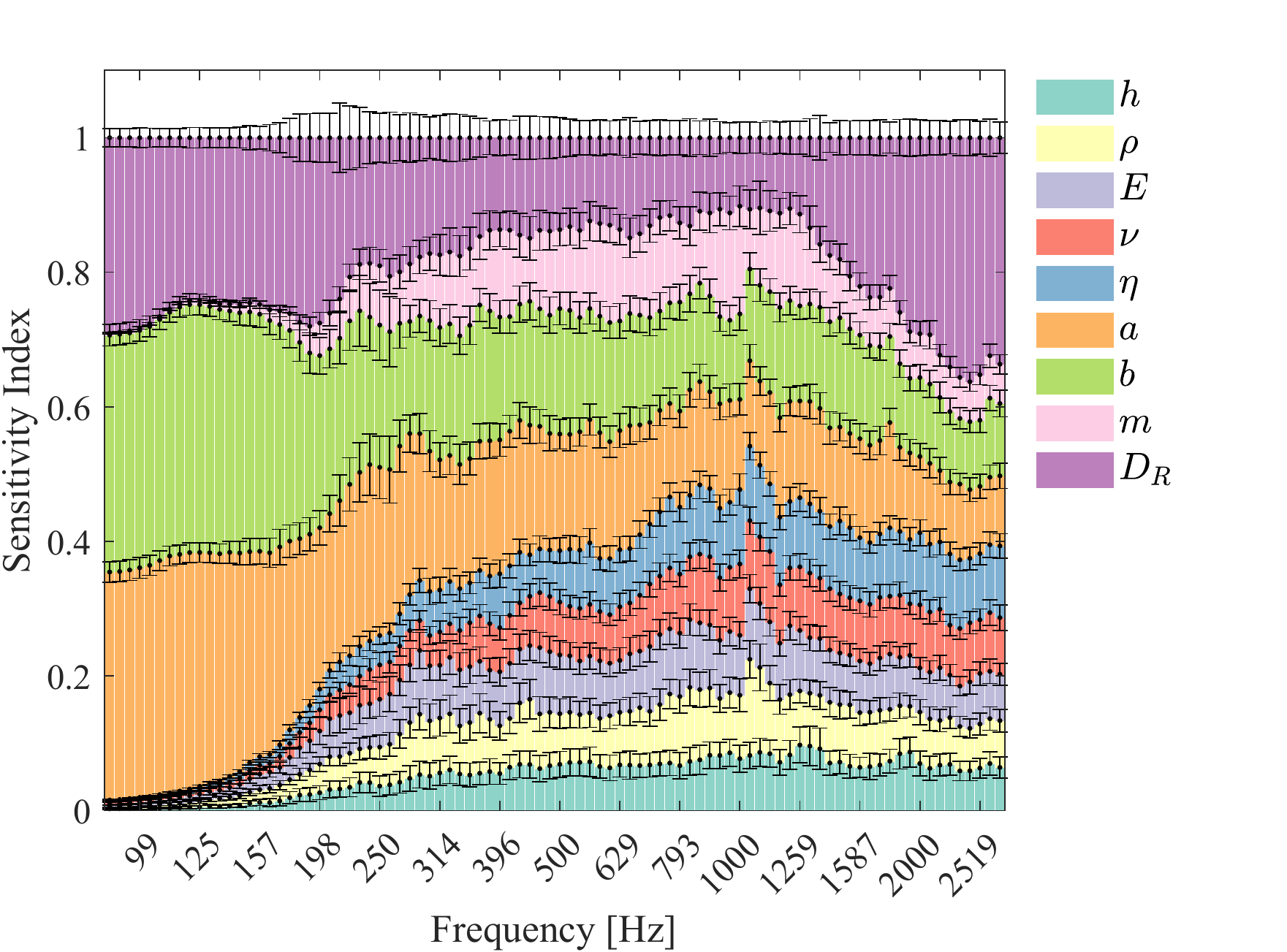}
         \caption{Physics-guided features $m$ and $D_R$ are included as input for the surrogate. $RMSE = 3.24  \pm 0.02 \ dB$. Training time = 24.86s} 
         \label{fig:FI_Modal_Summation_EI}
     \end{subfigure}
        \caption{MDI-based sensitivity of the STL of finite plates evaluated with the Modal Summation.} 
        \label{fig:FI_Modal_Summation}
\end{figure}

\begin{figure} [ht]
\captionsetup[subfigure]{justification=centering}
\centering 
     \begin{subfigure}[b]{0.37\textwidth}
         \centering
         \includegraphics[width=\textwidth]{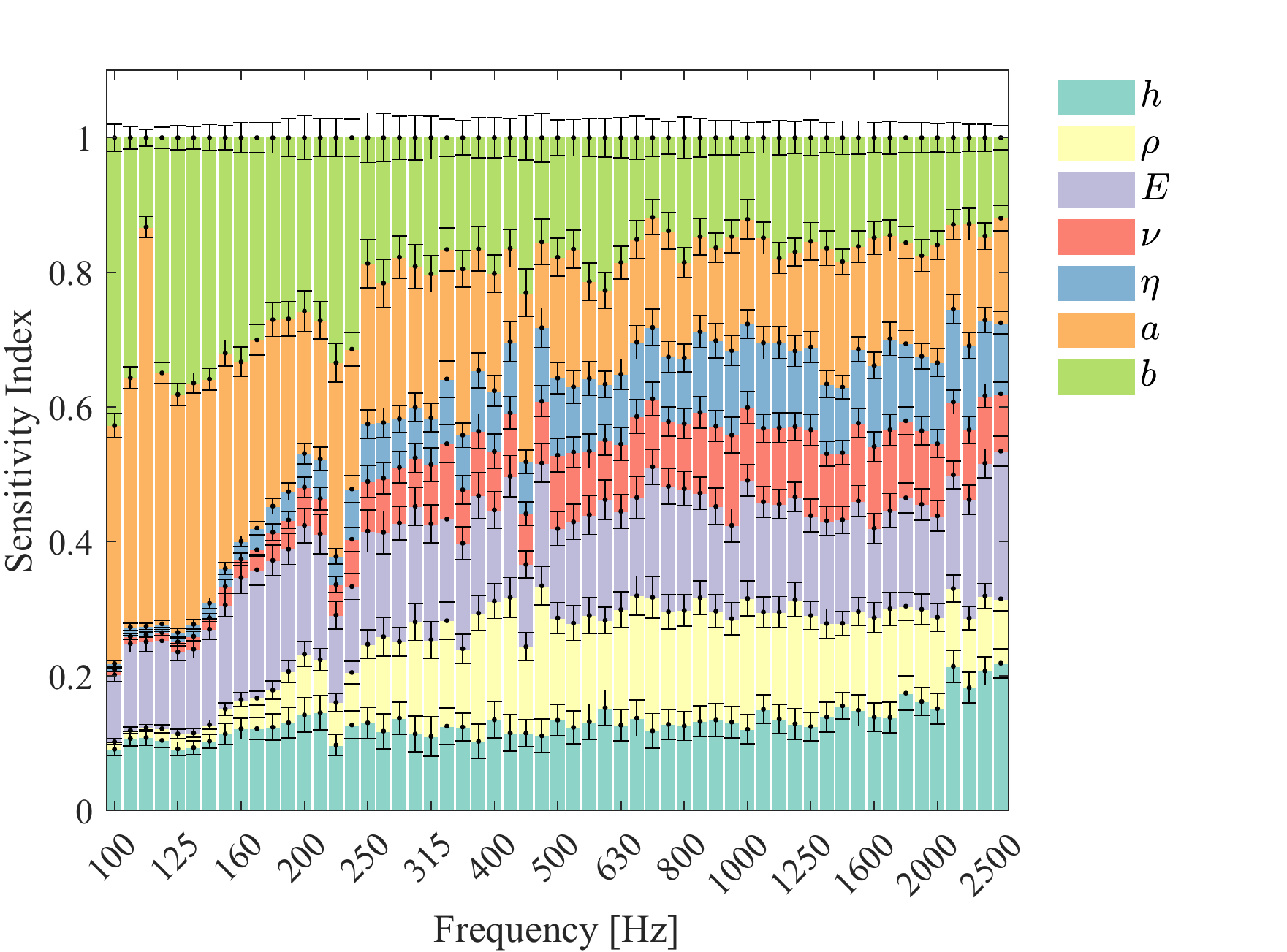}
         \caption{The inputs of the surrogate only consist of the STL variables. $RMSE = 5.76 \pm 0.06 \ dB$. Training time = 21.00s}
         \label{fig:FI_Comsol_BI}
     \end{subfigure}
     \hfill
     \begin{subfigure}[b]{0.37\textwidth}
         \centering
         \includegraphics[width=\textwidth]{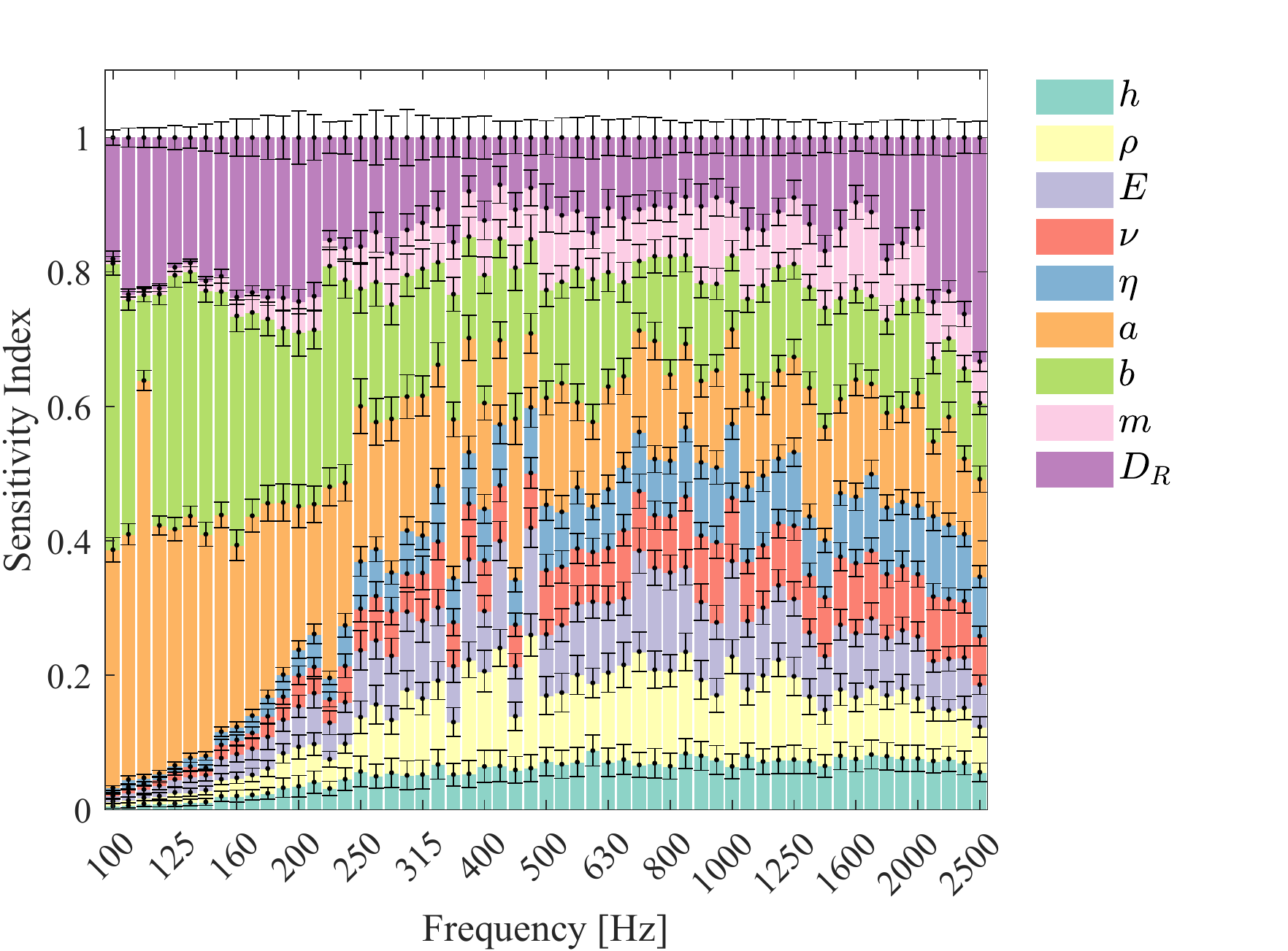}
         \caption{Physics-guided features $m$ and $D_R$ are included as input for the surrogate. $RMSE = 5.60  \pm 0.06 \ dB$. Training time = 21.16s}
         \label{fig:FI_Comsol_EI}
     \end{subfigure}
        \caption{MDI-based sensitivity of the STL of finite plates evaluated with FEM.} 
        \label{fig:FI_Comsol}
\end{figure} 

\begin{figure} [h]
\captionsetup[subfigure]{justification=centering}
\centering 
     \begin{subfigure}[b]{0.37\textwidth}
         \centering
         \includegraphics[width=\textwidth]{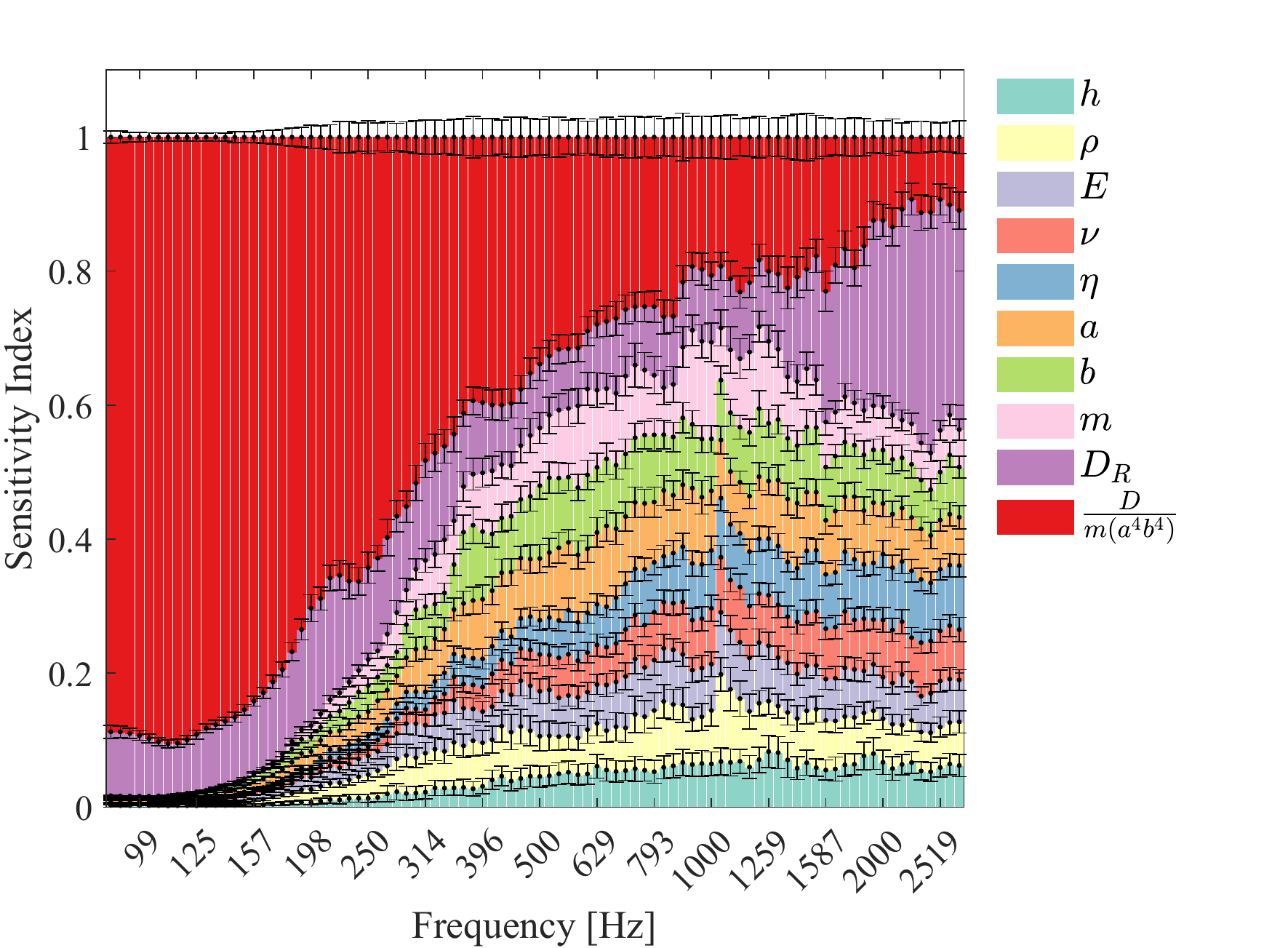}
         \caption{Sensitivity indices from Modal Summation dataset.$RMSE = 2.67 \pm 0.02 \ dB$. Training time = 32.62s}
         \label{fig:FI_MS_Resonant}
     \end{subfigure}
     \hfill
     \begin{subfigure}[b]{0.37\textwidth}
         \centering
         \includegraphics[width=\textwidth]{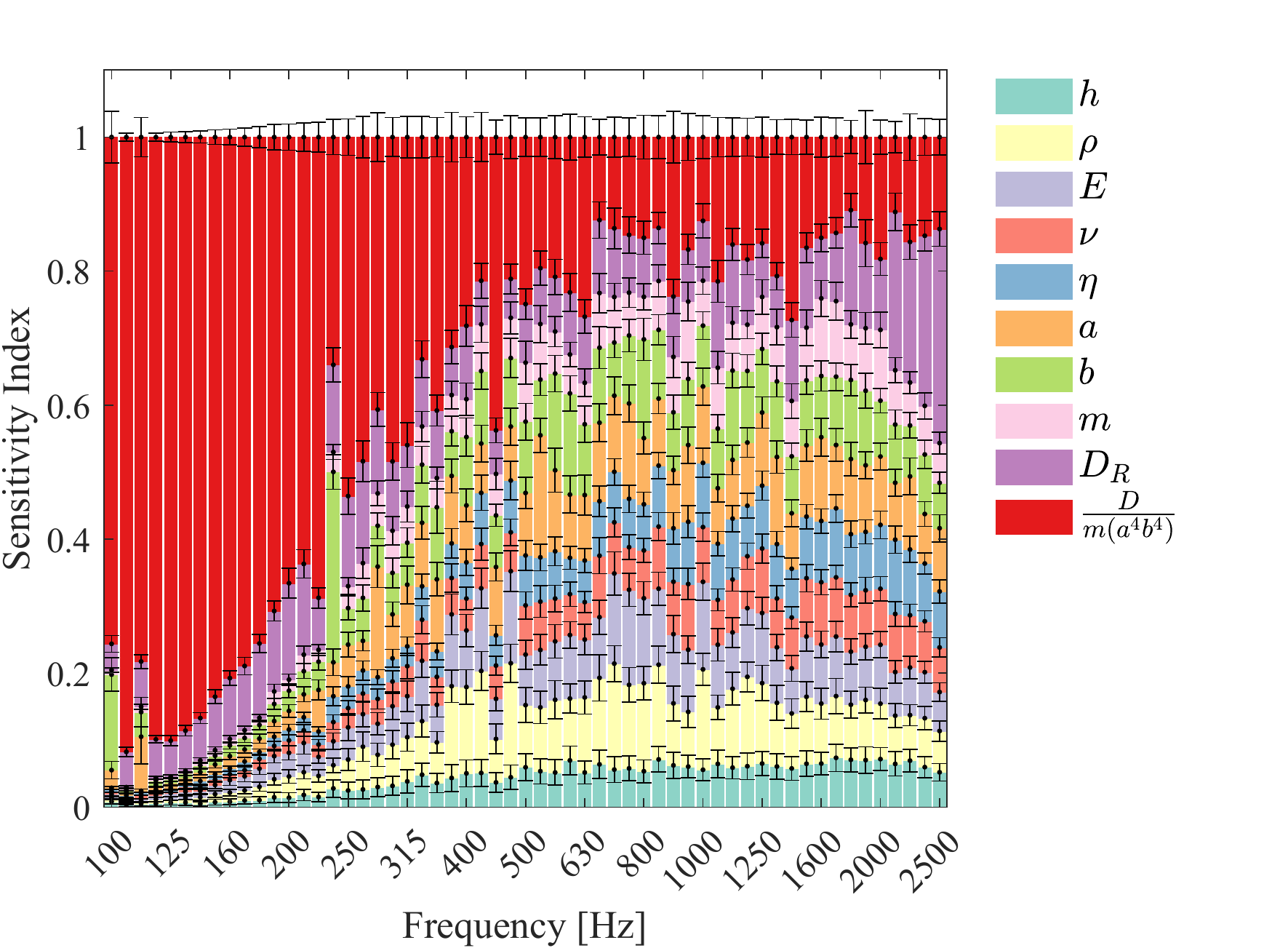}
         \caption{Sensitivity indices from FEM dataset. $RMSE = 4.86  \pm 0.05 \ dB$. Training time = 23.37s}
         \label{fig:FI_Comsol_Resonant}
     \end{subfigure}
        \caption{MDI-based sensitivity of the STL of finite plates including mass density $m$, the real part of bending stiffness $D_R$, and the resonance coefficient term $R$ as physics-guided features.} 
        \label{fig:FI_Resonant_Term}
\end{figure}

\begin{figure} [htp]
\captionsetup[subfigure]{justification=centering}
\centering 
     \begin{subfigure}[b]{0.37\textwidth}
         \centering
         \includegraphics[width=\textwidth]{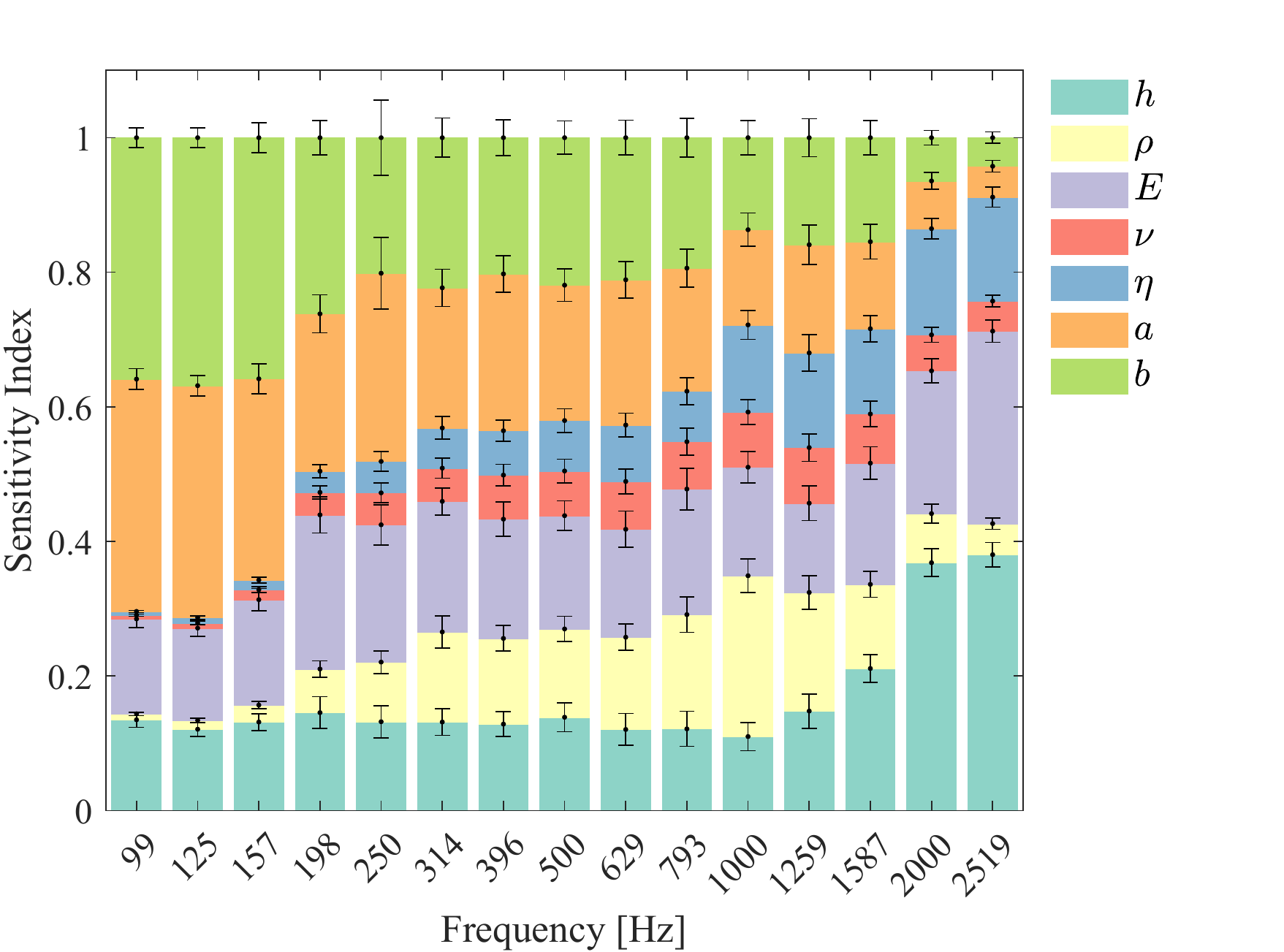}
         \caption{The inputs of the surrogate only consist of the STL variables. $RMSE = 2.19 \pm 0.05 \ dB$. Training time = 6.55s}
         \label{fig:FI_MS_avg_BI}
     \end{subfigure}
     \hfill
     \begin{subfigure}[b]{0.37\textwidth}
         \centering
         \includegraphics[width=\textwidth]{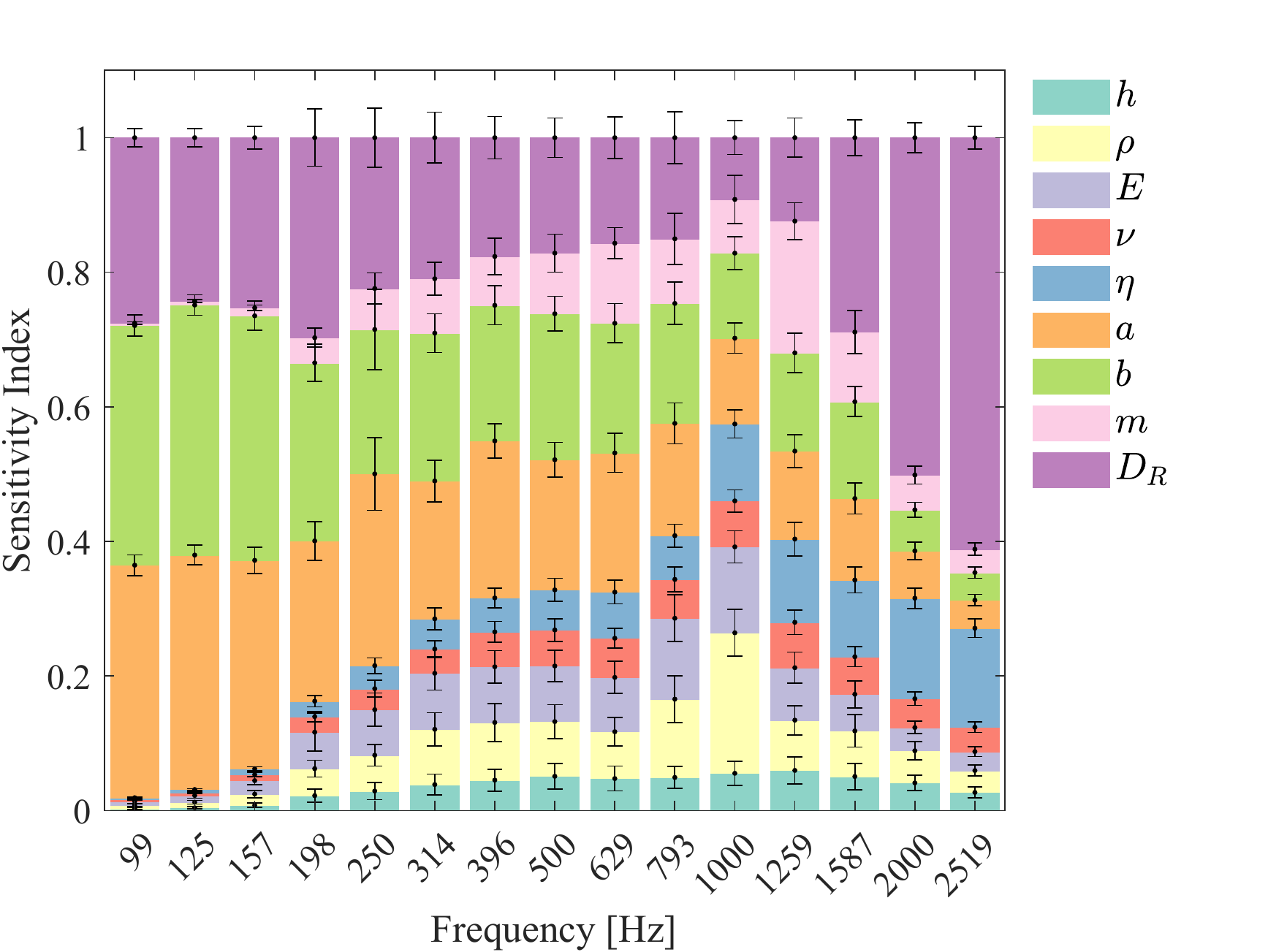}
         \caption{Physics-guided features $m$ and $D_R$ are included as input for the surrogate.  $RMSE = 2.01  \pm 0.02 \ dB$. Training time = 4.41s}
         \label{fig:FI_MS_avg_EI}
     \end{subfigure}
     \hfill
     \begin{subfigure}[b]{0.37\textwidth}
         \centering
         \includegraphics[width=\textwidth]{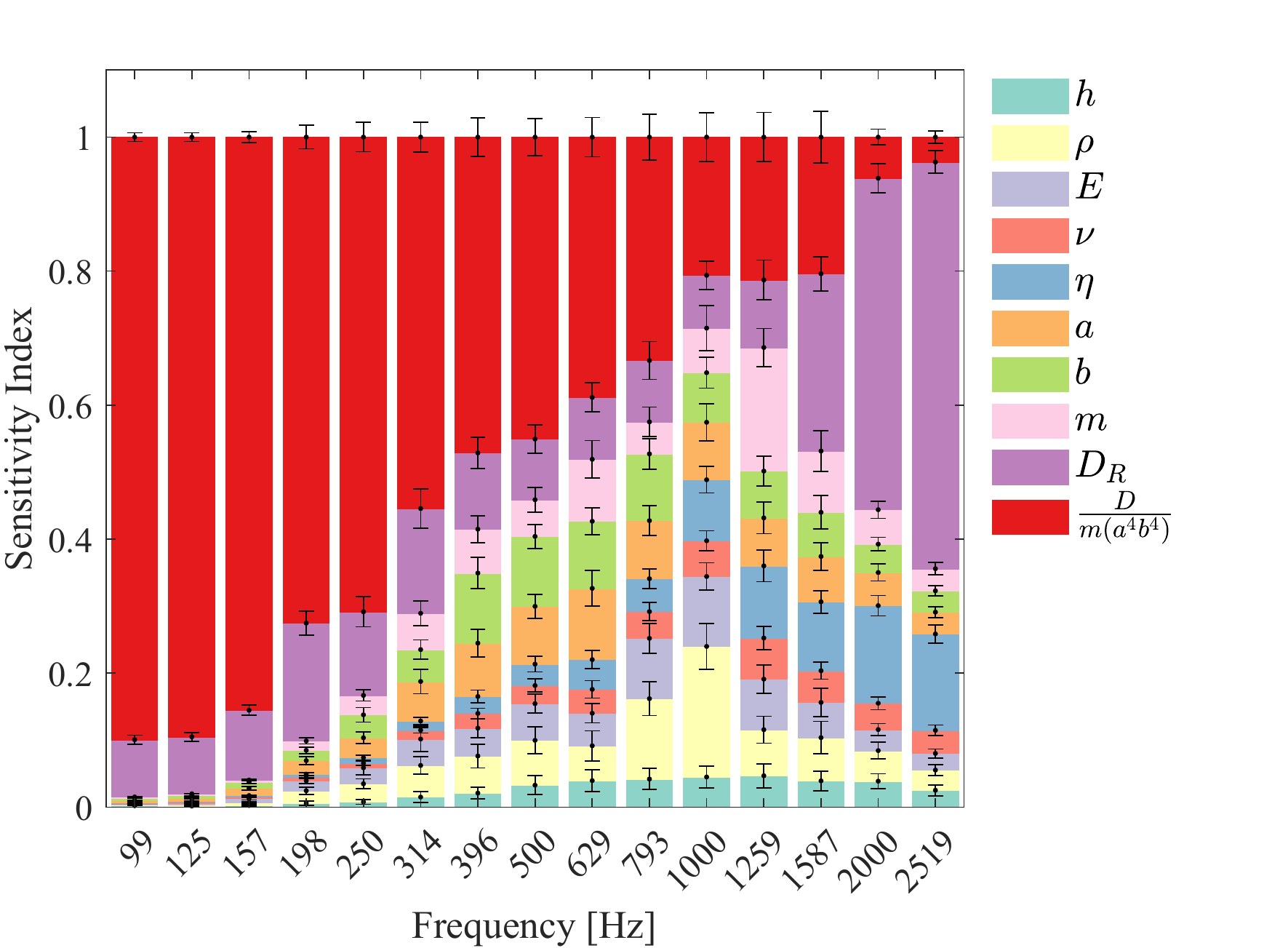}
         \caption{Physics-guided features $m$, $D_R$ and $R$ are included as input for the surrogate. $RMSE = 1.49  \pm 0.02 \ dB$. Training time = 5.22s}
         \label{fig:FI_MS_avg_Ress}
     \end{subfigure}
        \caption{MDI-based sensitivity of the one-third octave band average STL of finite plates evaluated with MS.} 
        \label{fig:FI_MS_avg}
\end{figure} 

\begin{figure}[htp]
\captionsetup[subfigure]{justification=centering}
\centering 
     \begin{subfigure}[b]{0.37\textwidth}
         \centering
         \includegraphics[width=\textwidth]{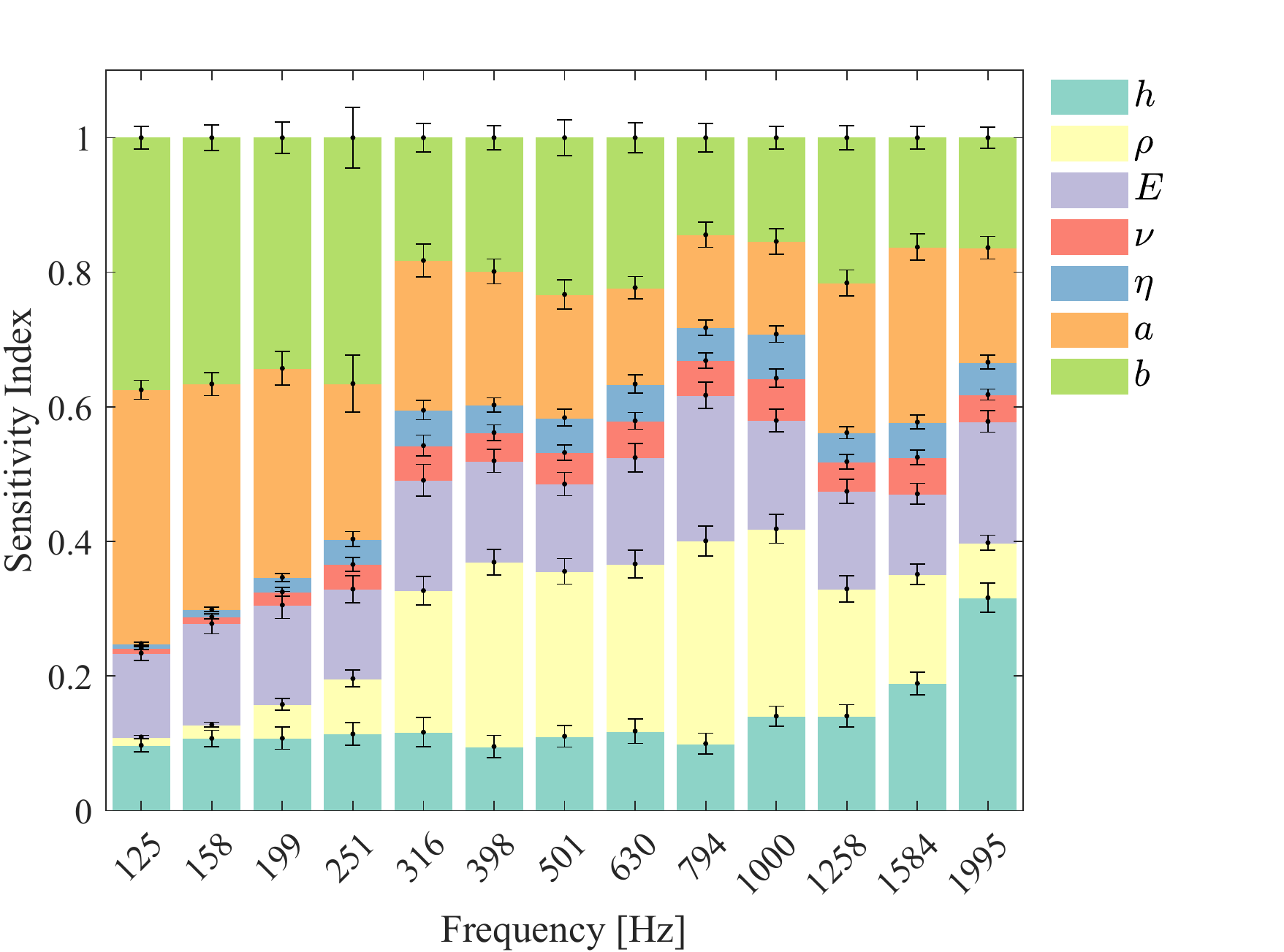}
         \caption{The inputs of the surrogate only consist of the STL variables. $RMSE = 2.23  \pm 0.02 \ dB$. Training time = 8.23s}
         \label{fig:FI_Comsol_avg_BI}
     \end{subfigure}
     \hfill
     \begin{subfigure}[b]{0.37\textwidth}
         \centering
         \includegraphics[width=\textwidth]{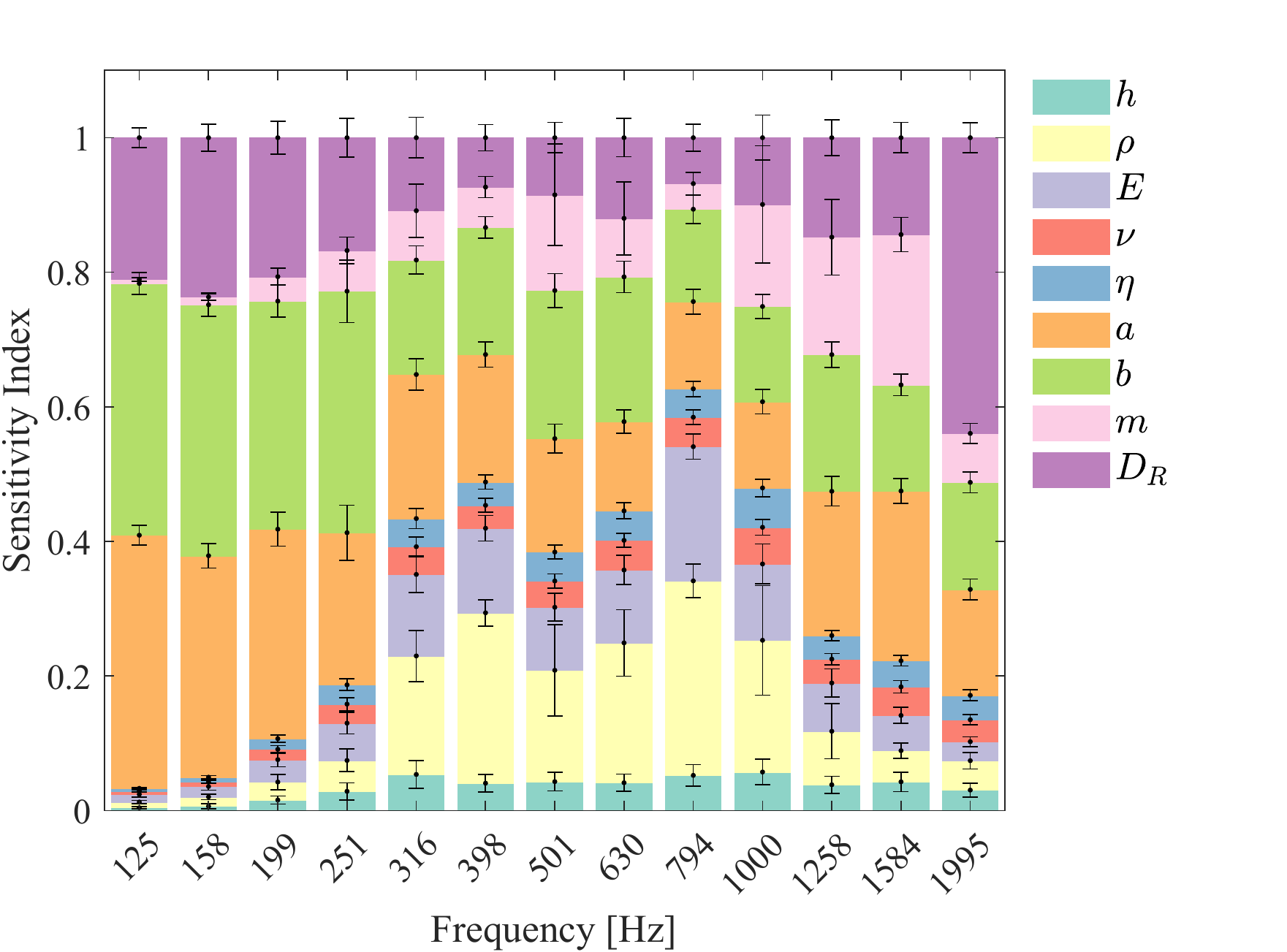}
         \caption{Physics-guided features $m$ and $D_R$ are included as input for the surrogate. $RMSE = 2.09  \pm 0.02 \ dB$. Training time = 6.34s}
         \label{fig:FI_Comsol_avg_EI}
     \end{subfigure}
     \hfill
     \begin{subfigure}[b]{0.37\textwidth}
         \centering
         \includegraphics[width=\textwidth]{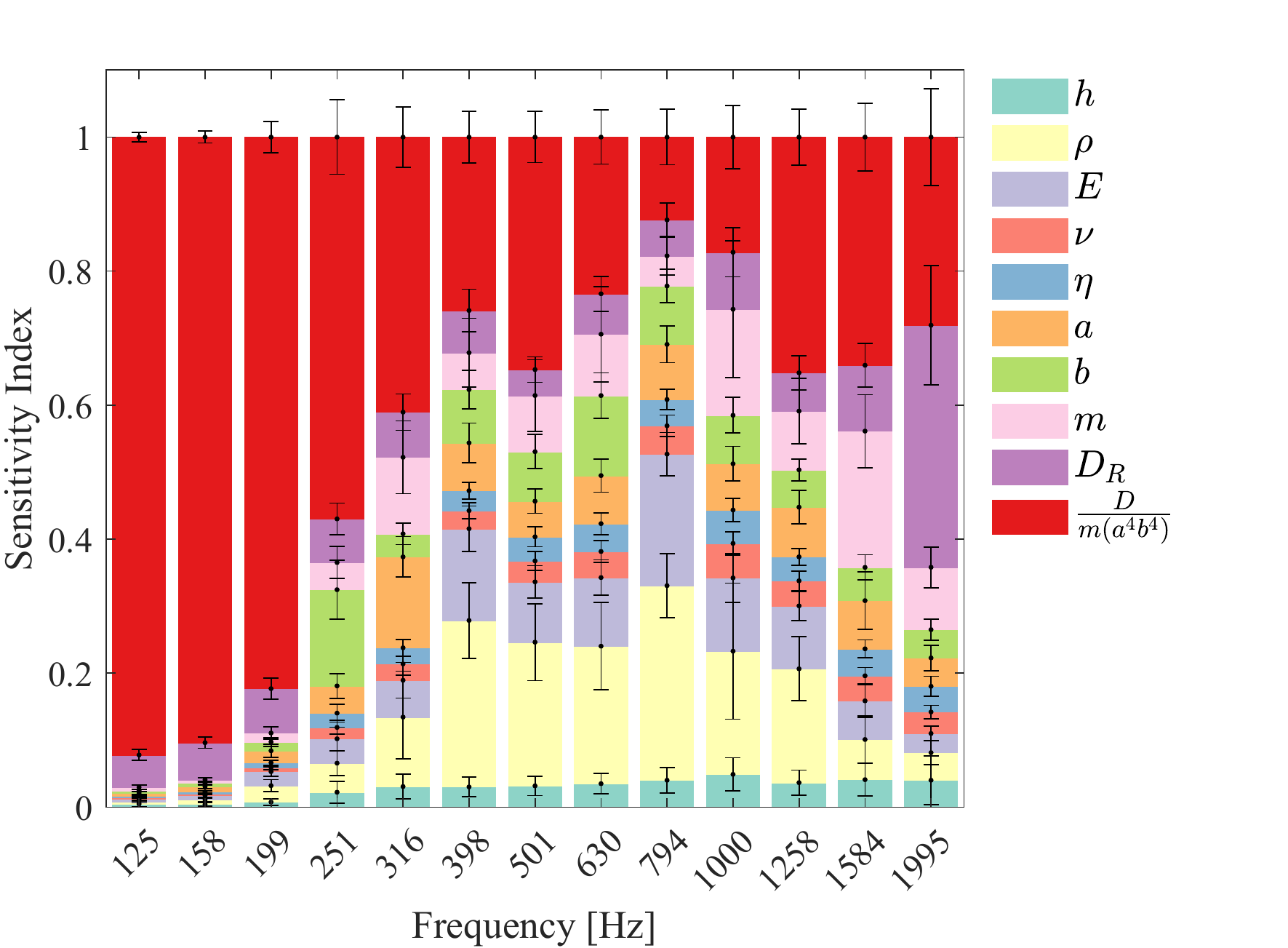}
         \caption{Physics-guided features $m$, $D_R$ and $R$ are included as input for the surrogate. $RMSE = 1.63  \pm 0.03 \ dB$. Training time = 5.01s}
         \label{fig:FI_Comsol_avg_Ress}
     \end{subfigure}
     
        \caption{MDI-based sensitivity of the one-third octave band average STL of finite plates evaluated with FEM.} 
        \label{fig:FI_Comsol_avg}
        
\end{figure} 

The MDI-based sensitivity indices considerably change when the model thoroughly considers the structural dynamics of the finite plate and its resonant modes, as for the STL results with MS and FEM approaches in Figures \ref{fig:FI_Modal_Summation} and \ref{fig:FI_Comsol}, respectively.
Overall, the sensitivity indices of MS and FEM approaches are similar to each other, indicating that they model the vibroacoustic phenomena with analogous considerations.
The bending stiffness and the plate dimensions have a considerable high sensitivity index throughout the frequency range.
Therefore, the behavior of the finite plates under analysis diverges from the one of a limp mass, an assumption held by the mass law and considered in the mass-controlled region of both the analytical approach for infinite plates and the correction factor approach for finite plates.
Indeed, in both Figures \ref{fig:FI_Modal_Summation} and \ref{fig:FI_Comsol}, the mass density importance is negligible in the resonant region, and its influence in the middle frequency range is not predominant.

The achieved surrogate accuracy with physics-guided features is $3.24 \pm 0.02 \ dB$ for the MS-based results and $5.60 \pm 0.06 \ dB$ for the FEM-based results. The errors are higher for the FEM model than for the MS model since the excitation in FEM is modeled by random components and, therefore, the results are non-deterministic.
The high complexity and non-smooth behavior of the STL response with ML and FEM approaches lead to poor surrogate accuracy even when the physics-guided features are used, which also compromises the accuracy of the sensitivity analyses.
In particular, at high frequencies, the even distribution of the sensitivity indices among the features indicates higher complexity and possibly higher inaccuracies.
As the modal behavior of the plate is posing difficulties for the surrogate model, the resonance coefficient term $R$ in \eqref{eqn:resonance_coef} is included as another physics-guided feature for the surrogates of the FEM and MS models. 

\begin{equation}
\label{eqn:resonance_coef}
    R = \frac{D}{m(a^4 b^4)}
\end{equation}

As observed in Figure \ref{fig:FI_Resonant_Term}, the resonance coefficient feature controls the STL in the low-frequency range. 
Although the accuracy improved to $2.67 \pm 0.02 \ dB$ for the MS-based results and to $4.86 \pm 0.05 \ dB$ for the FEM-based results, it remains elevated.
Extra improvements in the accuracy could be obtained by feature selection, that is, removing unimportant features, a technique commonly used in ML.

Finally, the MDI-based sensitivity index is evaluated with one-third octave band average STL with MS and FEM, as shown in Figures \ref{fig:FI_MS_avg} and \ref{fig:FI_Comsol_avg}, respectively. Once again, the stiffness and dimensions of the plate play a major role for all frequencies.
The surrogate models that include the three physics-guided features are the ones with the best accuracy, which is $RMSE = 1.49  \pm 0.02 \ dB$ for the MS model and $RMSE = 1.63  \pm 0.03 \ dB$ for FEM the model. 

Therefore, the results presented in this section show evidences of the physical consistency of the surrogate models as the STL sensitivity indices have a behavior coherent with the literature of each STL physics-driven model.
Furthermore, the sensitivity indices of the width and length of the plate are consistently equivalent in all analyses, as expected for isotropic materials.
Physical inconsistencies could indicate regions where the ML model is not accurate.
The investigation also demonstrates that physics-guided features can be readily included to improve surrogate accuracy and lead to a clearer distinction of STL regions in the sensitivity analysis.
However, the interpretation of the resulting feature importances is not straightforward when physics-guided features are included, as the input features are not independent anymore.
In summary, the MDI-based sensitivity analyses presented in this Section are a cheap-to-evaluate method that improves the surrogate interpretability and reliability and allows for the use of data-based information to deepen physical discussions on variable importance and interactions.

%%%%%%%%%%%%%%%%%%%%%%%%%%%%%%%%%%%%%%%%%%%%%%%%%%%%%%%%%%%%%%%%%%%%%%%%%%%%%%%%%%%
% Benchmarking
%%%%%%%%%%%%%%%%%%%%%%%%%%%%%%%%%%%%%%%%%%%%%%%%%%%%%%%%%%%%%%%%%%%%%%%%%%%%%%%%%%

\section{Benchmarking of ML Methods and Influence of the Design Space Selected}\label{sec:Benchmarking}

In this section, surrogate models based on NN, GPR, RF, and GBT are evaluated regarding accuracy and training time.
The benchmark is conducted for all STL models of Section \ref{sec:TL}.
The influence of physics-guided features is also considered in the test cases.
Supporting points are sampled from the design space of Table \ref{tab:SDS} using LHS.
Finally, an analysis of the influence of design space size and location in the surrogate accuracy is also performed.
For information about the models' hyperparameters, refer to~\ref{sec:sample:appendix2}.
It is worth mentioning that the configuration of the RF-based surrogate of this section differs from the one in Section \ref{sec:Feature Engineering}, as here just a single regressor predicts the outputs for all input frequencies.

\begin{figure} [ht]
\centering 
\includegraphics[width=0.48\textwidth]{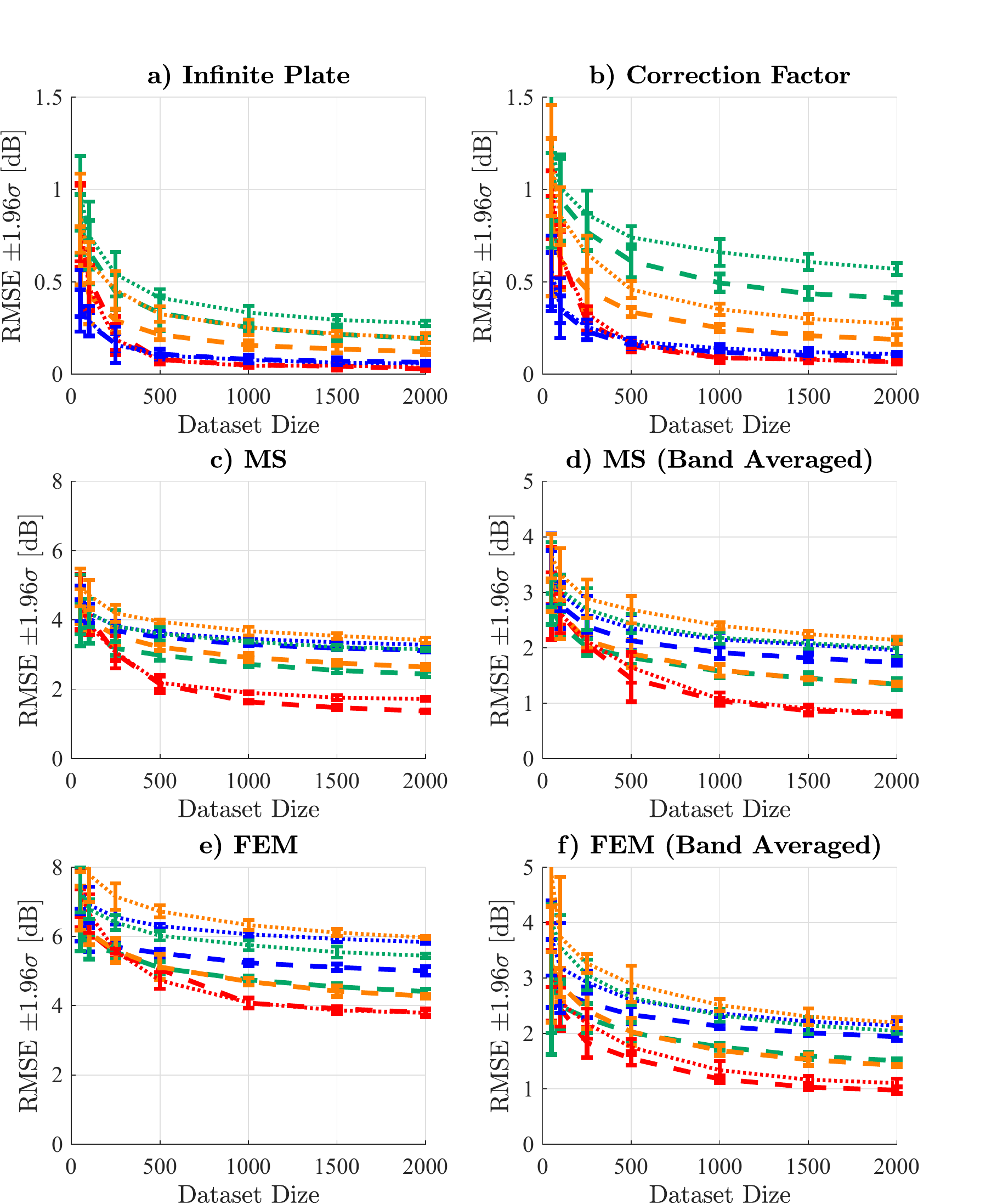} 
\caption{\label{fig:ErrorxPoints}
Benchmarking of Machine Learning-based surrogate regarding the accuracy in relation with the dataset size. {\tikzred{15pt}}~Neural Network, {\tikzblue{15pt}}~Gaussian Process Regressor, {\tikzgreen{15pt}}~Random Forest,  {\tikzpurple{16pt}}~Gradient Boosting Trees. With {\tikzdash{15pt}} and without {\tikzdot{13pt}} physics-guided features.}
\end{figure}

Figure \ref{fig:ErrorxPoints} shows the RMSE of each surrogate for all six STL models obtained with five-fold cross-validation.
The addition of physics-guided features consistently produces more accurate models. 
Moreover, it is observed that the NN-based surrogates yield better accuracy for all STL models.
This can be justified by the fact that NN can produce a better fit on non-smooth functions than other ML approaches \citep{imaizumi2019deep}.
However, as deep-learning demands big datasets, the NN does not have the best performance for small datasets.
Except by the FEM model, the NN-based surrogates of all STL models achieved RMSE below $3 \; \mathrm{dB}$, which are satisfactory results once measurements errors in STL experiments usually also go up to $3 \; \mathrm{dB}$.
The RMSEs for the FEM model are higher due to the stochastic excitation and overall more complex simulation.

\begin{figure} [h]
\centering 
\includegraphics[width=0.45\textwidth]{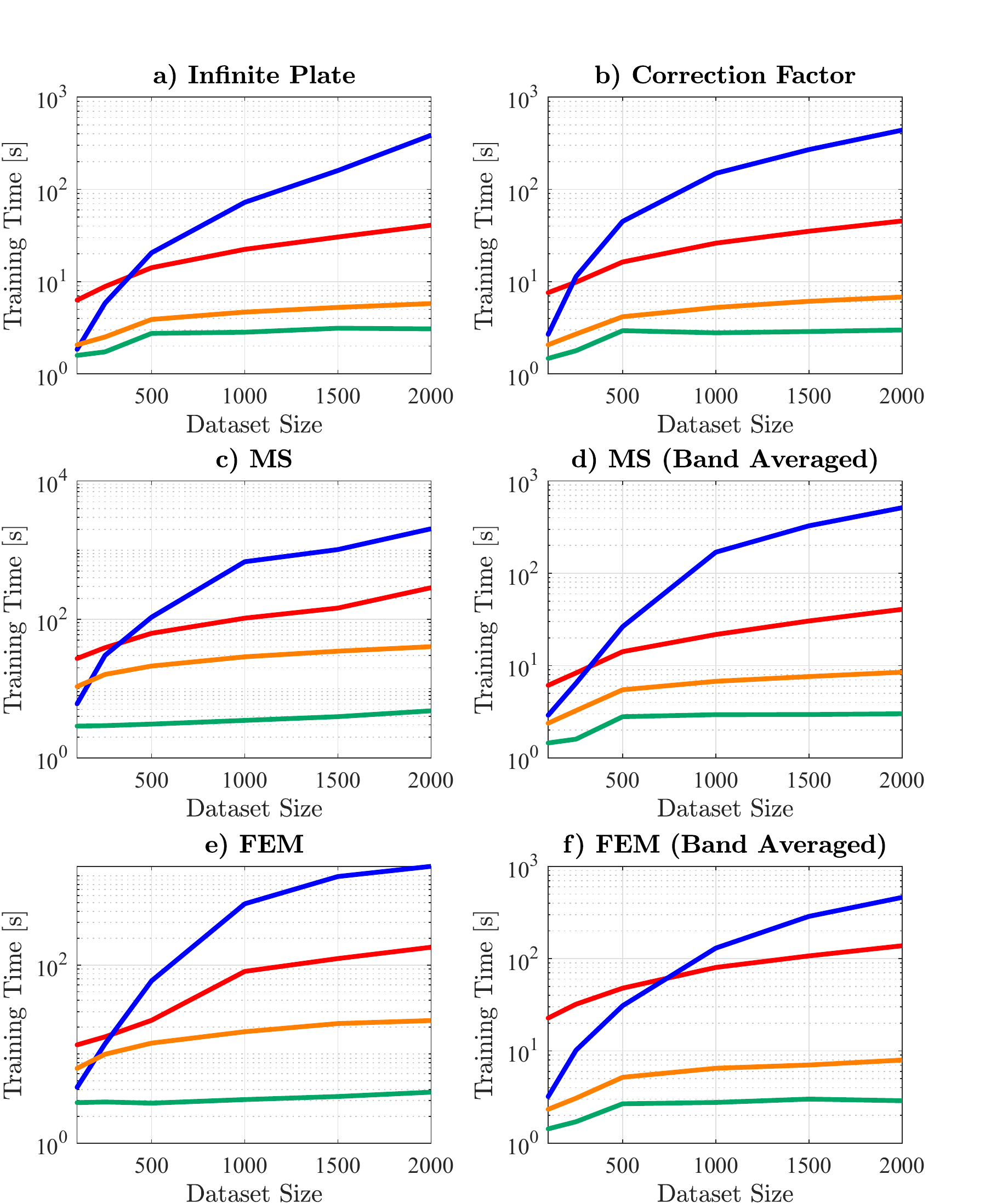} 
\caption{\label{fig:Time}Benchmarking of ML-based surrogate regarding the training time in relation with the dataset size. {\tikzred{15pt}}~Neural Network, {\tikzblue{15pt}}~Gaussian Process Regressor, {\tikzgreen{15pt}}~Random Forest,  {\tikzpurple{16pt}}~Gradient Boosting Trees.} 
\end{figure}

Another important factor in the choice of a surrogate model is the training time, which is presented in Figure \ref{fig:Time}.
It can be seen that RF and GBT surrogates are significantly faster to train than the other two methods.
This is due to the fact that the training of decision tree-based models scales loglinearly with the dataset size \cite{hastie2009elements}.
It is worth noting that, because no efficient multi-output implementation of the GBT model is available, it was required to fit one model for each input frequency, thus, increasing the overall training time compared to the RF model.
Further, GPR is computationally expensive to train for large datasets as its operations scales as $ \mathcal{O}(N^3)$ \citep{williams2006gaussian}.
%and NN relies on backpropagation for training the model parameters, which is computationally expensive \todo{citar}.
It is also worth noting that GPR, GBT, and RF have sensitivity analysis as a by-product and, therefore, are intrinsically more interpretable than NN. On the other side, NN and GPR extrapolate better than decision tree based methods.

\begin{figure} [h]
\centering 
\includegraphics[width=0.45\textwidth]{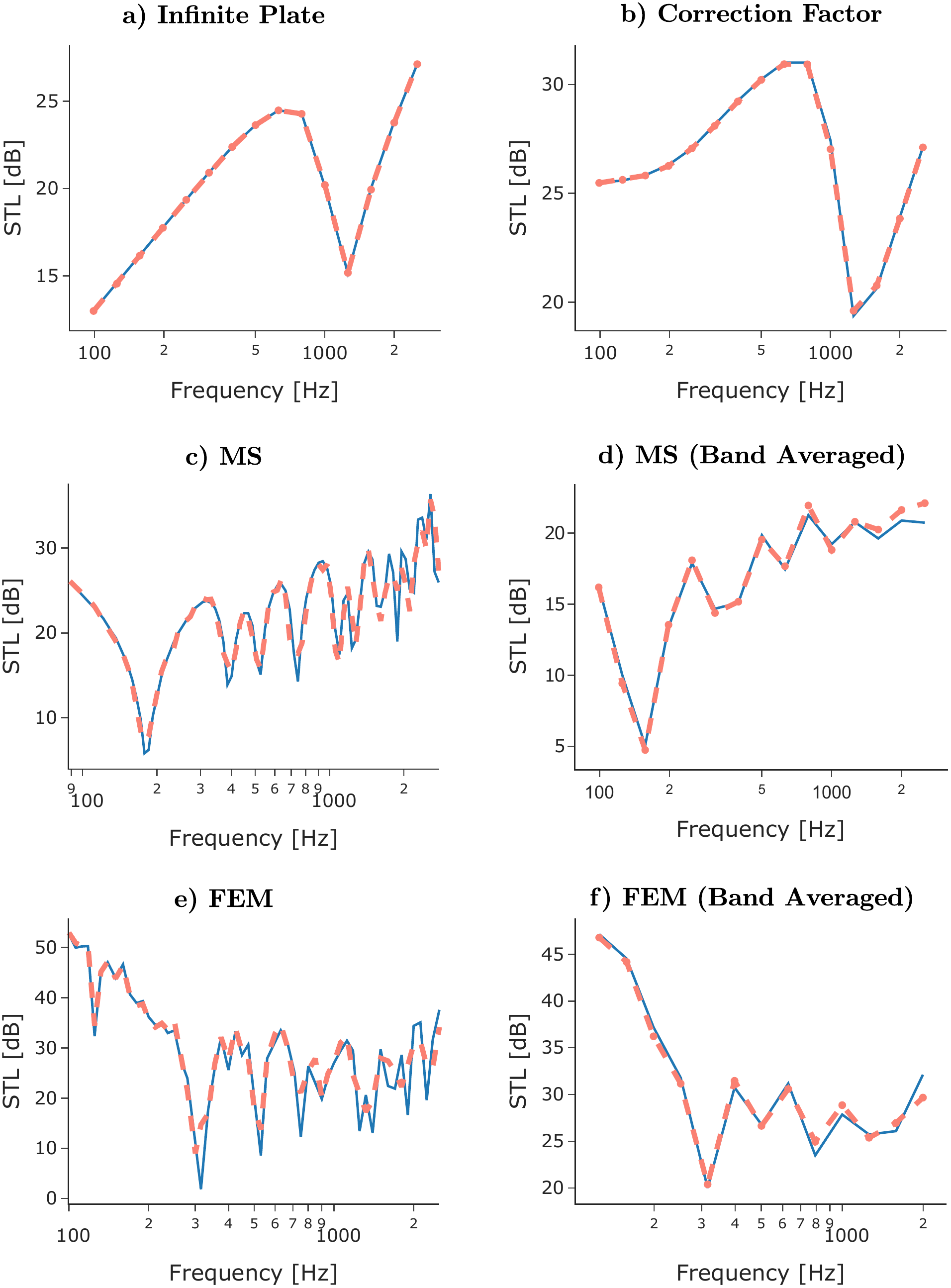} 
\caption{\label{fig:Predicted_Curves} Example of one predicted STL curve for each STL model using the NN-based surrogate with physics-guided features. {\tikzsim{15pt}}~Simulated, {\tikzpred{15pt}}~Predicted.} 
\end{figure}

\begin{figure} 
\centering 
\includegraphics[width=0.40\textwidth]{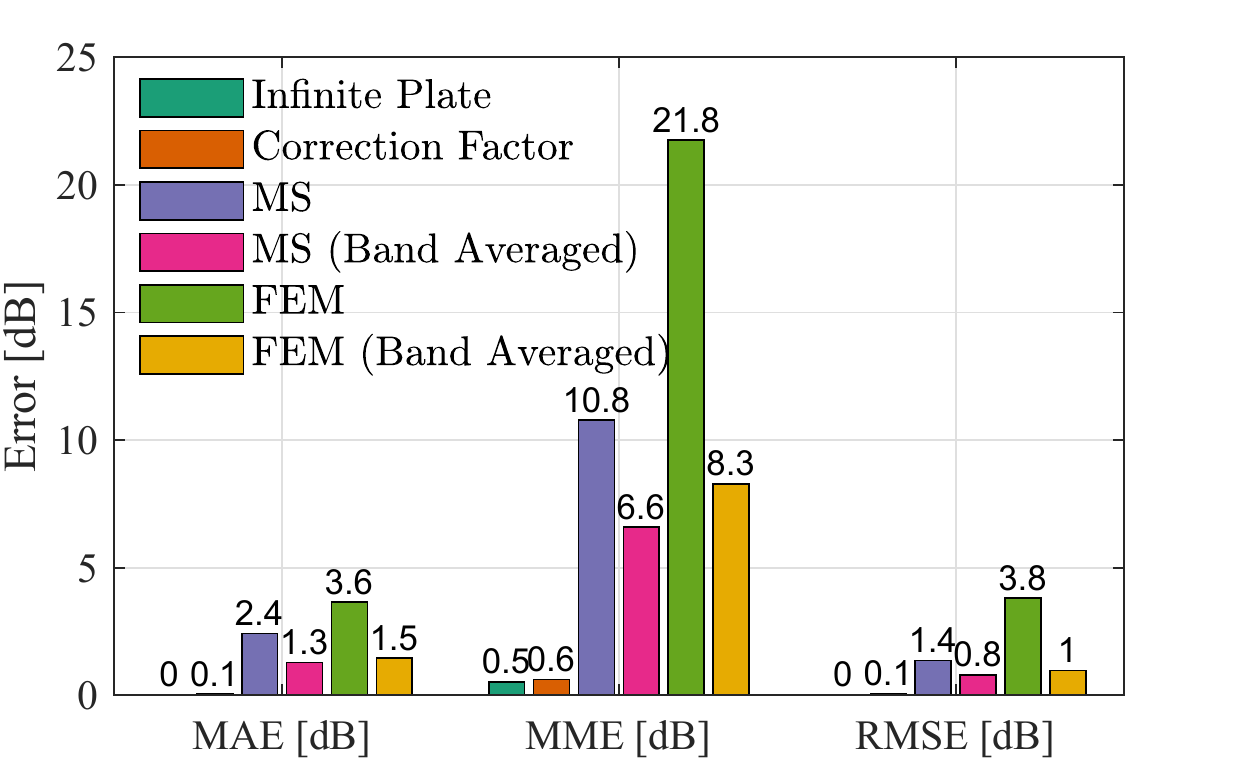} 
\caption{\label{fig:NN_errors}Mean absolute error (MAE), mean maximum error (MME) per STL design and root mean square error (RMSE) of the NN-based surrogate model with physics-guided features.} 
\end{figure} 

Figure \ref{fig:Predicted_Curves} shows an example of the predicted STL curve for each STL model using the most accurate surrogate, i.e. NN with physics-guided features trained over 2000 supporting points.
Visually, the predicted curves are remarkably similar to the simulated ones, apart from the surrogate trained for the FEM model.
Moreover, Figure \ref{fig:NN_errors} provides a deeper look into the NN-based surrogate model errors for each STL model, indicating additionally the mean absolute error (MAE) and the mean maximum error (MME).
For instance, the MME indicates that high localized errors occur in both MS and FEM models without band-average due to mispredicted local maxima and minima of the response signal.

%%%%%%%%%%%%%%%%%%%%%%%%%%%%%%%%%

\begin{figure*} [ht]
\centering 
\includegraphics[width=0.95\textwidth]{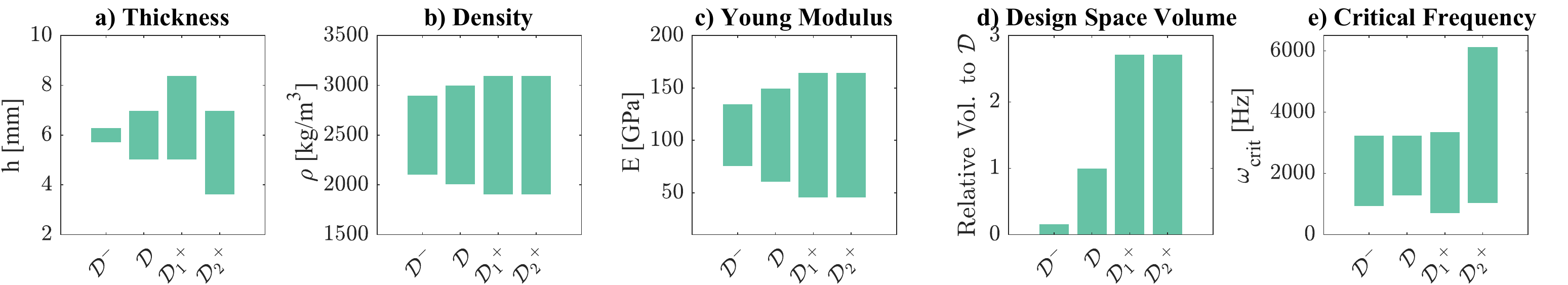} 
\caption{\label{fig:DS_Size_Loc} Range of the input variables used for each Design Space considered (a-c), the volume of the Design Spaces relative to the volume of the standard design space $ \mathcal{D}$ (d) and the range of critical frequencies resulting from each design space (e).} 
\end{figure*} 

An investigation on the NN-based surrogate performance regarding the set of interest $\hat{\mathcal{X}}$ was also performed. For that, four design spaces are considered, namely the standard design space $\mathcal{D}$ from Table \ref{tab:SDS}, the smallest design space $ \mathcal{D}^-$, and the biggest design space $ \mathcal{D}^+_1$ and $ \mathcal{D}^+_2$, which have the same size but are located in different places in the domain $ \mathcal{X}$, as illustrated in Figure \ref{fig:DS_Size_Loc}. 
The range of damping, Poisson’s ratio, and the width and length of the plates are the same from Table \ref{tab:SDS} for all design spaces. The resulting range of critical frequencies is also shown in Figure \ref{fig:DS_Size_Loc}.
For each design space, 500 supporting points were sampled using LHS for the STL models of the infinite plate, correction factor, and MS with and without one-third octave band average.

\begin{figure*} [ht]
\centering 
\includegraphics[width=0.9
\textwidth]{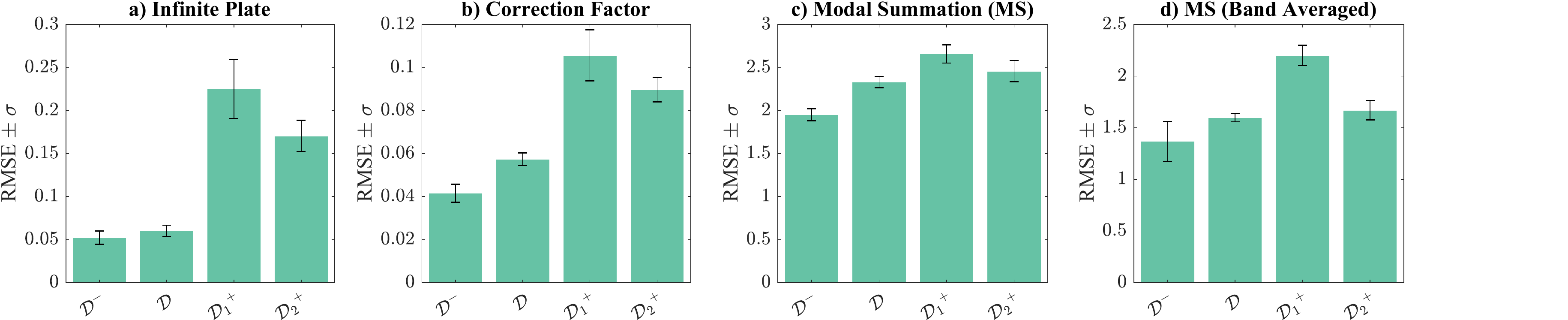} 
\caption{\label{fig:DS_Size_Results}NN-based surrogate accuracy for the design spaces $ \mathcal{D}$, $ \mathcal{D}^-$, $ \mathcal{D}^+_1$ and $ \mathcal{D}^+_2$ when modeling the STL problem using the model of infinite plates (a), correction factor model (b), Modal Summation model (c) and Modal summation with 1/3 octave band average.} 
\end{figure*}

Figure \ref{fig:DS_Size_Results} shows the accuracy obtained with an NN-based surrogate model trained with physics-guided features for each set of interest and STL method.
It is observed that the surrogate error increases with the design space size, especially for the infinite plate and correction factor models.
For the MS model without band average, the difference in the surrogates' accuracy is not relevant.

Furthermore, it can be observed that a considerable difference in accuracy is obtained for design spaces $ \mathcal{D}^+_1$ and $ \mathcal{D}^+_2$, although they have the same size.
This is because the samples from $ \mathcal{D}^+_2$ have, on average, a simpler behavior than the samples from $ \mathcal{D}^+_1$, once many of them present critical frequency higher than 2500 Hz which is above the frequency range modeled by the surrogates.

%%%%%%%%%%%%%%%%%%%%

In conclusion, the developed surrogate models present overall satisfactory performance for most of the studied cases.
Even highly discontinuous and rough functions could be properly predicted, which is an essential property for surrogates of vibroacoustic simulations.
Moreover, in particular for STL models, NN-based surrogates are the most accurate, and, thus, are useful tools to explore the design space of STL simulations, enabling optimal and robust vibroacoustic designs.
It has also been observed that the location of the design space can have more of an effect on surrogate accuracy than its size.

%%%%%%%%%%%%%%%%%%%%%%%%%%%%%%%%%%%%%%%%%%%%%%%%
%%%%%%%%%%%%%%%%%%%%%%%%%%%%%%%%%%%%%%%%%%%%%%%%
% Conclusion
%%%%%%%%%%%%%%%%%%%%%%%%%%%%%%%%%%%%%%%%%%%%%%%%
%%%%%%%%%%%%%%%%%%%%%%%%%%%%%%%%%%%%%%%%%%%%%%%%

\section{Discussion and Conclusion}\label{sec:Conclusions}

The presented study investigates the use of Machine-Learning (ML)-based surrogate models to overcome the computational burden of expensive vibroacoustic analyses.
It is shown how to tackle the lack of interpretability of black-box simulators and what are the surrogates' capacities to fit the non-smooth functions which are recurrent in the vibroacoustic domain.
As a study case, surrogate models are constructed for Sound Transmission Loss (STL) analyses modeled at different complexities levels, including analytical and numerical approaches. 

The physical consistency of the surrogate models was verified through sensitivity analyses intrinsic to the ML methods, without additional computational cost.
For each region of the frequency spectrum, the physical behavior identified by the features' importance results were in accordance with the STL theory.
Moreover, aspects such as the differences between the STL models were inferred from the sensitivity study, providing a deeper understanding of the modeled phenomena.

The benchmarking of surrogates based on four ML methods, namely Neural Networks (NN), Gaussian Process Regressor (GPR), Random Forest (RF), and Gradient Boosted Trees (GBT) show that NN consistently performs better.
The surrogate presented satisfactory accuracy even for functions with highly non-smooth behavior resulting from resonant and coincidence phenomena.
Furthermore, it is shown that including domain knowledge through physics-guided features improves the surrogate performance for all ML methods and STL models.
The surrogate errors are also shown to be proportional to the design space size for the surrogate with less complex STL models but do not change significantly with the design space size for the STL with high non-smooth responses. On the other side, the location of the design space in the input domain has a relevant impact on the accuracy in all cases.

Although the results were generated for the case of STL, some conclusions can be generalized to other vibroacoustic problems which have similar behavior. First, the inclusion of physics-guided features can be readily applied to other physical domains, as well as the sensitivity analysis methods. Second, the aptitude of the ML-based surrogate, especially NN-driven, to predict non-smooth behavior was witnessed in the case of STL and is an indicator for other vibroacoustic analyses. Therefore, this paper presents good practices and methods to create ML-based surrogate models for vibroacoustic problems with improved interpretability, accuracy, and physical guidance. The results show the potential of using the surrogate as an informed decision-making tool to make domain exploration, uncertainty quantification, and optimization viable.

%%%%%%%%%%%%%%%%%%%%%%%%%%%%%%%%%%%%%%%%%%%%%%%%
% Acknowledgements
%%%%%%%%%%%%%%%%%%%%%%%%%%%%%%%%%%%%%%%%%%%%%%%%

\section{Acknowledgements}\label{sec:Acknowledgements}

This work received financial support of the European Union’s Horizon 2020 research and innovation program under Marie-Curie grant agreement No 860243 to the LIVE-I project.

%%%%%%%%%%%%%%%%%%%%%%%%%%%%%%%%%%%%%%%%%%%%%%%%
% Declaration of Competing Interest
%%%%%%%%%%%%%%%%%%%%%%%%%%%%%%%%%%%%%%%%%%%%%%%%

\section{Declaration of Competing Interest}\label{sec:Declaration_Interest}

The authors declare that they have no known competing financial interests or personal relationships that could have appeared to influence the work reported in this paper.

%%%%%%%%%%%%%%%%%%%%%%%%%%%%%%%%%%%%%%%%%%%%%%%%
% Appendix
%%%%%%%%%%%%%%%%%%%%%%%%%%%%%%%%%%%%%%%%%%%%%%%%

\appendix

\section{Methodology of the STL Analyses}
\label{sec:sample:appendix1}

The methodologies used to model the STL analyses are described in this appendix, following the notation in Figure \ref{fig: STL}.

The plate is in the plane $z=0$, separating the source field $P|_{z>0}$ and the transmission field $P|_{z<0}$, both filled with a light fluid with density $\rho_0$ and characteristic sound speed $c_0$. Considering that an harmonic plane wave with wavenumber vector

\begin{align}
\label{eqn:wave_vector}
    \bm{k} &= {\begin{pmatrix} k_x & k_y & k_z \end{pmatrix}}^T \\ \nonumber 
    &=  \frac{\omega}{c_0}{\begin{pmatrix} \sin{\theta}\cos{\phi} & \sin{\theta}sin{\phi} & \cos{\theta} \end{pmatrix}}^T 
\end{align}

\noindent impinges the plate with incident angle $\theta$ and azimuth angle $\phi$, the resultant incident acoustic pressure field is

\begin{align}
    P|_{z>0} &=  P_I + P_R \\ \nonumber 
    &= p_I e^{i(\omega t- k_xx-k_yy-k_zz)} +  p_R e^{i(\omega t-k_xx-k_yy+k_zz)},
\end{align}

\noindent where $p_I$ is the amplitudes of the incident wave in the incident field $P_I$, and $p_R$ is the amplitude of the wave reflected by the plate, which forms the reflected field $P_R$.
The plate vibration also radiates a plane wave of magnitude $p_T$ in the transmission field $P_T$, which acoustic pressure is given by:

\begin{equation}
    P|_{z<0} = P_T = p_T e^{i(\omega t- k_xx-k_yy-k_zz)}.
\end{equation}

The structural-acoustic coupling equation states that there must be continuity of acoustic and mechanics velocities at the interface of the plate with the acoustic fields:

\begin{equation}
    \frac{\partial{P}}{\partial{z}} \Bigr|_{z=0_\pm} = {\rho_0 \omega^2 w},
\end{equation}

\noindent where $w = W(x,y,z)e^{(i \omega t)}$ is the plate transverse displacement. For the sake of brevity, the factor $e^{(i \omega t)}$ will be omitted from the equations hereafter. Therefore, the coupling equations in the source and transmission side are, respectively:

\begin{equation}
\label{eqn:coupling_source}
     i k_z(P_R-P_I) = {\rho_0 \omega^2 w},
\end{equation}
\begin{equation}
\label{eqn:coupling}
    -i k_z P_T = {\rho_0 \omega^2 w},
\end{equation}

\noindent from what the following relationship between the pressure fields amplitudes is obtained:

\begin{equation}
\label{eqn:pr}
    P_R = P_I-P_T.
\end{equation}

By definition, the acoustic transparency $\tau$ is the ratio between transmitted and incident sound intensity, where the sound intensity can be written as:

\begin{equation}
\label{eqn:Sound_Intensity}
    I_i = |P_i\cdot v_i| = \frac{|P_i|^2}{\rho_0 c_0},
\end{equation}

\noindent where $v_i$ is the local fluid velocity. Therefore, the acoustic transparency is:

\begin{equation}
\label{eqn:tau}
    \tau(\phi, \theta,\omega) =  \left|\frac{P_T}{P_I}\right|^2.
\end{equation}

The acoustic transparency for a diffuse field $\tau_d(\omega)$ is the weighted average transmission coefficient of all possible incident and azimuth angles:

\begin{equation}
\label{eqn:tau_diff}
    \tau_{d}(\omega)= \frac{\int_{0}^{2\pi} \int_{0}^{\pi/2} \tau(\phi, \theta,\omega) cos{\theta}sin{\theta} d\theta d\phi}
    {\int_{0}^{2\pi} \int_{0}^{\pi/2} cos{\theta}sin{\theta} d\theta  d\phi }.
\end{equation}

The STL is the inverse of the acoustic transparency measured on the dB scale.

To find the relationship between the pressure field amplitudes in Eq. \ref{eqn:tau}, the constitutive equation of motion of the plate should be solved combined with Eq. \ref{eqn:pr}, which for an isotropic plate can be written as:

\begin{align}
\label{eqn:isotropic_plate}
D\nabla^4{w}-\omega^2 m w &=  P|_{z=0^-} - P|_{z=0^+} \\ \nonumber
&=  (P_I-P_R)- (P_T) \\ \nonumber
&=  2(P_I-P_T). \nonumber
\end{align}

The solution for each approach addressed in this paper is presented below.

%%%%%%%%%%%%%%%%%%%%%%%%%%%%%%%%%%%%%%%%%%%%%%%%

\bigskip

\textbf{\emph{Analytical model of infinite plates.}} For the case of a infinite an isotropic plate, the plate response is independent of $\phi$, therefore:

\begin{equation}
    \nabla = \frac{\partial}{\partial{\mathbf{x}}} = -i\mathbf{k} = -i \frac{\omega}{c_0} sin{\theta},
\end{equation}

\noindent and Eq. \ref{eqn:isotropic_plate} is rewrite as:

\begin{equation}
\label{eqn:iso2}
    \left( D\frac{\omega^4}{c_0^4} sin^4{\theta} - m \omega^2 \right) w = 2(P_I-P_T).
\end{equation}

Thus, the structural impedance of the plate reads as:

\begin{equation}
\label{eqn:isotropic_impedance}
    Z(\omega,\theta) = \left(1 - \frac{\omega^2 D}{m c_0^4}sin^4{\theta} \right) i \omega m.
\end{equation}

Combining Eq. \ref{eqn:coupling}, \ref{eqn:iso2} and \ref{eqn:isotropic_impedance}, it is possible to find the relation:

\begin{equation}
    \frac{P_T}{P_I} =  \left(\frac{Z cos{\theta}}{2 \rho_0 c_0} +1 \right)^{-1},
\end{equation}

\noindent which is straightforward applied in Eq \ref{eqn:tau} to find the acoustic transparency for a plane wave. As Eq. \ref{eqn:tau_diff} does not have a close analytical solution, it is numerically evaluated to output STL for a diffuse field.

%%%%%%%%%%%%%%%%%%%%%%%%%%%%%%%%%%%%%%%%%%%%%%%%

\bigskip

\textbf{\emph{Correction Factor.}} 
Based in the Rayleigh-integral method, \citep{atalla2006modeling} proposes to apply a correction factor to the acoustic transparency of the infinite plate $\tau_{\infty}$ to account for effects of the plate size, such that the acoustic transparency of a finite plate could be approximated by:

\begin{equation}
    \tau_{fin} =  (\sigma_R cos\theta) \tau_{\infty},
\end{equation}

\noindent where the radiation efficiency $\sigma_R $ is given by

\begin{equation}
    \sigma_R(\theta,\phi,a,b) =  \frac{\Re (Z_{fin}(\theta,\phi,a,b))}{\rho_0 c_0}.
\end{equation}

The structural impedance $Z_{fin}$ is evaluated as:

\begin{align}
    Z_{fin} = \frac{j\rho_0\omega}{S} \int_S\int_S 
    e^{-i(k_xx+k_yy)}G(x,y;x',y') \\  \nonumber
    \times \ e^{-i(k_xx'+k_yy')}dxdydx'dy',
\end{align}

\noindent  $S$ designates the surface of the panel and $G$ is the half-space Green's function. \citet{rhazi2010simple} proposes analytical simplifications, so that $\sigma_R$ simplifies into:

\begin{align}
    \sigma_R(\theta,\phi,a,b) = \frac{\omega b}{16c_0}\int_{-1}^1 \frac{a}{b} F(\theta,\mu(\phi),a,b) d\mu,
\end{align}

\noindent where $\mu = (4\phi/\pi-1)$. This equation can be evaluated using Gaussian numerical integration. The function $F(\theta,\mu(\phi),a,b)$ and the details of its derivation are found in \citep{rhazi2010simple}.

%%%%%%%%%%%%%%%%%%%%%%%%%%%%%%%%%%%%%%%%%%%%%%%%

\bigskip

\textbf{\emph{Modal Summation.}}
An analytical solution of the fluid-structure iteration problem of a simple supported plate can also be obtained by modeling the plate displacement $w$ in terms of a modal summation:

\begin{align}
    \label{eqn:modal_sum_w}
    w(x,y,z) = \sum_{m,n}\alpha_{m,n}\varphi_{m,n}(x,y),
\end{align}

\noindent where $\sum_{m,n}$ is the short format for $\sum^{\infty}_{m=1}\sum^{\infty}_{n=1}$, the subindices $m$ and $n$ indicate the mode index, $\alpha_{m,n}$ is the coefficient of contribution of each mode for the displacement field, and $\varphi_{m,n}(x,y)$ (Eq. \ref{eqn:mode_shape}) is the modal function that satisfy the simple supported boundary conditions (Eq. \ref{eqn:BC}).

\begin{align}
\label{eqn:mode_shape}
    \varphi_{m,n}(x,y) = sin\left(\frac{m \pi x}{a}\right) sin\left(\frac{n \pi y}{b}\right).
\end{align}

\begin{align}
\label{eqn:BC}
    w = \frac{\partial^2w}{\partial x^2} + \nu \frac{\partial^2w}{\partial y^2} = 0,&\quad\text{for  }x = \pm a/2, \\ \nonumber
    w = \frac{\partial^2w}{\partial y^2} + \nu \frac{\partial^2w}{\partial x^2} = 0,&\quad\text{for  }y = \pm b/2.
\end{align}

Using the 2D Fourier Transform, each pressure field $P_{i}$ can also be described as a modal summation:

\begin{align}
\label{eqn:modal_sum_p}
P_{i}(x,y,z) = \sum_{m,n}{p_{i}}_{m,n}\varphi_{m,n}(x,y)e^{\pm k_z z},
\end{align}

\noindent where the signal of the exponential function depends on the propagation direction of the plane wave in the field, and in which the coefficient of contribution of the pressure field ${p_{i}}_{m,n}$ read as:

\begin{align}
\label{eqn:p_mn}
    {p_{i}}_{m,n} = \frac{4}{ab} \int^b_0 \int^a_0 p_i e^{-i(k_xx+k_yy)}\varphi_{m,n}(x,y) dx dy.
\end{align}

Substituting Eq. \ref{eqn:modal_sum_w}, \ref{eqn:modal_sum_p} and \ref{eqn:p_mn} into Eq. \ref{eqn:isotropic_plate} and applying the weighted residuals method:

\begin{dmath}
\label{eqn:Galerkin}
    \int_0^b \int_0^a
    \left[ D \nabla^4
    \left( \sum_{m,n} \alpha_{m,n} \varphi_{m,n} \right) - \omega^2  m \sum_{m,n} \alpha_{m,n} \varphi_{m,n} 
    - \sum_{m,n} 2\left({p_{I}}_{m,n} - {p_{T}}_{m,n}\right) \varphi_{m,n} 
    \right] \varphi_{m,n} dx dy = 0.
\end{dmath}

As the modal functions have orthogonal properties $\{\varphi_{j,k}\}^T M \{\varphi_{h,g}\}=0$ for $j\ne h$ or $k\ne g$, Eq. \ref{eqn:Galerkin} reads as

\begin{dmath}
\label{eqn:Galerkin2}
    \alpha_{m,n} D  \frac{\int_0^b \int_0^a \left[ \nabla^4 \varphi_{m,n} \right] \varphi_{m,n} dx dy }
    {\int_0^b \int_0^a \varphi_{m,n} dx dy }
    -  \alpha_{m,n} \omega^2 m  - 2 \left( {p_{I}}_{m,n} - {p_{T}}_{m,n}\right) = 0
\end{dmath}

\noindent if inter-modal coupling is neglected. The integrals on the first term can be analytically evaluated and the relation from Eq. \ref{eqn:coupling_source} is used to result, resulting in:

\begin{align}
\label{eqn:Galerkin3}
    \left[
    \omega^2_{m,n} -  \omega^2 + 
    2 \frac{i \omega \rho_0 c_0 }{m cos\theta}
    \right] \alpha_{m,n} = \frac{2}{m} {p_{I}}_{m,n},
\end{align}

\noindent where $\omega^2_{m,n}$ are the natural frequencies of the simply supported plate and ${p_{I}}_{m,n}$ can be analytically evaluated by solving the integrals from Eq \ref{eqn:p_mn}. Finally, the solution of the system of equation in Eq. \ref{eqn:Galerkin3} provides the coefficients of modal participation $\alpha_{m,n}$, and, consequently, the displacement $w$ (Eq. \ref{eqn:modal_sum_w}). The coefficients ${p_{T}}_{m,n}$ are evaluated with Eq. \ref{eqn:coupling} and the acoustic transparency is obtained by:

\begin{align}
\label{eqn:tau_MS}
    \tau(\theta,\phi)=\frac{ \sum_{m,n}|{p_{T}}_{m,n}|^2}{  \sum_{m,n}|{p_{I}}_{m,n}|^2}.
\end{align}

%%%%%%%%%%%%%%%%%%%%%%%%%%%%%%%%%%%%%%%%%%%%%%%%

\bigskip

\textbf{\emph{Finite Element Method.}} The numerical solution of the STL problem was implemented in COMSOL Multiphysics\textsuperscript{\textregistered} software \citep{COMSOL_soft} using FEM. The plate was modeled by shell elements with simply supported boundary conditions, while the receiver acoustic field was modeled by 3D finite elements with Perfect Matched Layer (PML) in its external interfaces to simulate the free-field condition. The FEM model does not include the acoustic field on the source side, but an equivalent pressure load is applied to the plate to account for the fluid load. The fluid pressure is modeled as the summation of 200 components in random directions to simulate the diffuse field. The frequency-domain formulation is used to solve the coupled plate constitutive equation and acoustic wave equation. The incident and transmitted power are then calculated as the integral of the incident and transmitted intensity on the plate surface, respectively. Details on the theory and numerical implementation can be found in \citep{COMSOL_guide}.

%%%%%%%%%%%%%%%%%%%%%%%%%%%%%%%%%%%%%%%%%%%%%%%%
% Hyperparameters
%%%%%%%%%%%%%%%%%%%%%%%%%%%%%%%%%%%%%%%%%%%%%%%%

\section{Machine-Learning Configuration for Benchmarking Analysis}
\label{sec:sample:appendix2}

The hyperparameters of the ML methods used in Section \ref{sec:Benchmarking} are detailed here. In all cases, the test size was 20\% of the dataset.

%%%%%%%%%%%%%%%%%%%%%%%%%%%%%%%%%%%%%%%%%%%%%%%%

\bigskip

\textbf{\emph{NN-based surrogate models.}} The NN are implemented using \textit{Keras} with \textit{tensorflow} as backend. One regressor is trained to output all STL predictions at $n_{outputs}$ frequencies. Standardization is applied to all input features, while the outputs are scaled in the range [0, 1]. The fully connected NNs have an architecture with $n_{inputs}$ nodes in the input layer, five hidden layers with 32 nodes each, and $n_{outputs}$ nodes in the output layer, where $n_{inputs}$ is the number of input features, which depends in the STL model and in the use of engineered features. All hidden layers have the \textit{sigmoid} as activation function and a L2 regularization penalty of $1e^{-7}$. The training was performed with batches of 32 samples and with 1500 epochs, except for MS and FEM models, in which the NN trained for 2500 epochs. The Adam algorithm was applied to minimize the average mean squared error of the STL curve prediction.

%%%%%%%%%%%%%%%%%%%%%%%%%%%%%%%%%%%%%%%%%%%%%%%%

\bigskip

\textbf{\emph{GPR-based surrogate models.}} The kernel $k$ used to construct the GPR uses a constant kernel $k_C$, a Matern 3/2 kernel $k_M$, a Radial Basis Function kernel $k_{RBF}$ and a white noise kernel $k_{\sigma}$ combined as $k(x_i,x_j) = k_C(x_i,x_j)*k_M(x_i,x_j) + k_{RBF}(x_i,x_j) + k_{\sigma}(x_i,x_j)$. The kernel hyperparameters are optimized to maximize the log-marginal likelihood of the posterior distribution using the L-BFGS-B algorithm with ten restarts. One gaussian process regressor predicts the STL at $n_{outputs}$ frequencies. The preprocessing of the inputs and outputs is the same from the NN-based surrogate models. 

%%%%%%%%%%%%%%%%%%%%%%%%%%%%%%%%%%%%%%%%%%%%%%%%

\bigskip

\textbf{\emph{RF-based surrogate models.}} An RF was trained to output the STL for the entire frequency range analyzed. The RF is made up of 200 CARTs without a depth limit. The impurity criterion is the mean of the averaged squared error among the $n_{outputs}$ outputs.

%%%%%%%%%%%%%%%%%%%%%%%%%%%%%%%%%%%%%%%%%%%%%%%%

\bigskip

\textbf{\emph{GBT-based surrogate models.}} The GBT predictor is made up of 125 decision trees with a maximum depth of 10 nodes. The learning rate of the training was set to 0.05. A different GBT predictor is trained for each predicted output, willing to minimize the mean squared error.

%%%%%%%%%%%%%%%%%%%%%%%%%%%%%%%%%%%%%%%%%%%%%%%%
%%%%%%%%%%%%%%%%%%%%%%%%%%%%%%%%%%%%%%%%%%%%%%%%
% Bibliography
%%%%%%%%%%%%%%%%%%%%%%%%%%%%%%%%%%%%%%%%%%%%%%%%
%%%%%%%%%%%%%%%%%%%%%%%%%%%%%%%%%%%%%%%%%%%%%%%%
\bibliographystyle{elsarticle-num-names} 
\bibliography{cas-refs}

\end{document}